\documentclass[aps,preprint,tightenlines,floatfix,groupedaddress]{revtex4}

\usepackage{graphicx}
\usepackage{dcolumn}

\begin{document}
% You should use BibTeX and apsrev.bst for references
\bibliographystyle{apsrev}

%\preprint{\texttt hep-ph/01xxxxx}

\title{Ubiquitous CP violation in a top-inspired left-right model}
% Optional argument for running titles on pages
%\title[]{}


% other informatio
\author{Ken Kiers}
\email{knkiers@tayloru.edu}
\author{Jeff Kolb}
\email{jeff_kolb@tayloru.edu}
\author{John Lee}
\email{john_lee@tayloru.edu}
%\thanks{}
\affiliation{Physics Department, Taylor University, 236 West Reade Ave.,
Upland, Indiana 46989}
\author{Amarjit Soni}
\email{soni@bnl.gov}
\affiliation{High Energy Theory, Department of Physics, Brookhaven
National Laboratory, Upton, New York 11973-5203}
\author{Guo-Hong Wu}
\email{gwu@darkwing.uoregon.edu}
\affiliation{Institute of Theoretical Science, University of Oregon, Eugene,
Oregon 97403-5203}

\date{\today}

\begin{abstract}
We explore CP violation in a Left-Right Model that reproduces the quark mass
and CKM rotation angle hierarchies in a relatively natural way by fixing
the bidoublet Higgs VEVs to be in the ratio $m_b:m_t$.  Our model is quite
general and allows
for CP to be broken by both the Higgs VEVs and the Yukawa couplings.
Despite this generality, CP violation may be parameterized in terms of two basic phases.
A very interesting feature of the model is that the mixing angles in the right-handed sector
are found to be equal to their left-handed counterparts to a very good approximation.
Furthermore, the right-handed analogue of the usual CKM phase $\delta_L$ is found to 
satisfy the relation $\delta_R \approx \delta_L$.
The parameter space of the model is explored by using an
adaptive Monte Carlo algorithm and the allowed regions in parameter
space are determined by enforcing experimental constraints from the $K$ and $B$ systems.  
This method of solution allows us to evaluate the left- and right-handed CKM matrices numerically 
for various combinations of the two fundamental CP-odd phases in the model.  
We find that all experimental constraints may be satisfied with right-handed $W$
and Flavour Changing Neutral Higgs masses as low as about 2~TeV and 7~TeV,
respectively.
\end{abstract}

%\pacs{}

\maketitle

\section{Introduction}
\label{intro}

The left-handedness of the observed weak interactions has long been a source
of curiosity in particle physics.  Left-right symmetry may be restored to the
weak interactions at the Lagrangian level by introducing a new right-handed
gauge boson.  The aesthetic appeal of the so-called Left-Right Model
has led many to study it over the past few decades
and the formal properties of the model are well-known~\cite{patisalam,mohapatrapati,senjanovic,fritzsch,
olness,gunion,deshpande,cho,reina,bgnr,rodriguez}.  One feature that has emerged is that
the new right-handed gauge boson must have a mass
in the TeV range in order to evade the stringent
bounds imposed by $\Delta m_K$~\cite{bbs}.  This mass scale
was unattainable two decades ago, but is now within reach, especially at upcoming
colliders and perhaps also through precision studies of low-energy observables in the $B$ system. 
Another factor that motivates a re-examination of the Left-Right Model
is that the model naturally accomodates non-vanishing neutrino masses
as well as the enormous disparity in 
masses observed among the quarks and leptons.  Indeed, very light neutrinos may be obtained
through the See-Saw Mechanism, while the heaviness of the top quark may be reproduced
through a judicious choice of Vacuum Expectation Values (VEVs) in the
extended Higgs sector of the model.  The model is also able to account for the observed
CP violation in the kaon and $B$ systems and has additional possibilities for non-standard CP
violation through the presence of extra CP-odd phases.  These phases appear, for example,
in the right-handed analogue of the usual 
Cabibbo-Kobayashi-Maskawa (CKM) matrix~\cite{cabibbo,km} and are in addition to the
single phase that appears in the usual CKM matrix.

While the formal properties of the Left-Right Model are well-known, its parameter
space has not been studied exhaustively except in certain limiting cases.  Two such
cases are represented by the quasimanifest and pseudomanifest versions of
the model.  In the former, CP violation is present explicitly in the Yukawa couplings;
in the latter, it arises spontaneously in the 
Higgs VEVs.  In both of these cases the right-handed analogue
of the usual CKM matrix is simplified
in that the three right-handed rotation angles are identical to their left-handed counterparts.
In the nonmanifest version of the model (considered in the present work), CP violation occurs
in both the Higgs VEVs and the Yukawa couplings, and the right-handed CKM matrix
can in principle be quite different from the left-handed one.  In this case a full numerical
solution needs to be undertaken in order to obtain detailed 
information regarding the right-handed sector of the model.

Detailed numerical results were first obtained for the Left-Right Model in the early 1980s, mostly
within the context of the pseudomanifest version of the model~\cite{ecker80,ecker81,ecker85,ecker86}.
The authors of Ref.~\onlinecite{frere} 
improved upon earlier approximate methods of solution,
while those of Refs.~\onlinecite{ball1} and \onlinecite{ball2} imposed combined
constraints coming from the neutral $K$ and $B$ systems.  These latter works
were all performed within the context of the pseudomanifest version of the model.
To our knowledge, a detailed numerical solution of the 
nonmanifest case (CP violation in the Higgs VEVs and in the Yukawa couplings), such as we present here, has not been performed.
One important consideration in any numerical treatment of the Left-Right Model concerns the
Flavour Changing Neutral Higgs (FCNH) that is generically present.
The FCNH contribution to $\epsilon_K$ occurs at tree level, leading in principle to a
prohibitively large (of order 50~TeV~\cite{pospelov}) lower bound on the Higgs mass scale.  
Our numerical study indicates that significantly lower values for the Higgs mass 
-- on the order of 7 TeV -- are actually
tolerable.    	

In this paper we undertake a relatively exhaustive search of the parameter space
of the Left-Right Model, while making few assumptions regarding the structure
of the model.  Two main features distinguish the present work from that of previous
authors.  In the first place, we allow for CP violation both in the Yukawa couplings and in the
Higgs VEVs (hence the ``ubiquitous'' in the title of this paper~\footnote{When considering CP violation
in the Higgs VEVs,
we focus on the bidoublet Higgs and ignore the right-handed triplet, since the latter does not
affect the quark mass matrices.  See, for example, Ref.~\onlinecite{rodriguez}.}).  
In the second, we employ a novel approach to the numerical solution of the problem,
using a Monte Carlo algorithm to search
the parameter space of the model.
Our main assumption concerns the extended Higgs sector, where we take
the bidoublet Higgs VEVs to be in the ratio $m_b:m_t$.  
This assumption is quite appropriate in the Left-Right Model and leads very
naturally to the observed hierarchy in the left-handed CKM matrix~\cite{ecker80,ecker81,ecker85,frere}.
We also show that this assumption leads naturally to other attractive features of this model, namely that
the rotation angles in the right-handed CKM matrix equal their left-handed counterparts to a good approximation
and that the CKM phases $\delta_R$ and $\delta_L$ are approximately equal
(see Appendix~\ref{sec:appendixb}).
Note that while we shall always fix the ratio of the bidoublet Higgs VEVs, 
our method of solution is quite powerful and could easily be generalized
to the case where the ratio is not $m_b:m_t$.

Throughout the present work we stress an important and general result that may not be widely known:
assuming a minimal Higgs sector and three generations of quarks, the
quark mass matrices in the Left-Right Model depend on at most two non-removable phases.
This insight allows for the numerical
solution of what might otherwise be a very complicated problem.
Perhaps more importantly, the
model contains only one new CP-odd degree of freedom beyond the one in the Standard Model (SM),
a very desirable feature when comparing the model to forthcoming precision experimental results (particularly
those coming from the $B$ factories).
In our notation, one of the CP-odd phases comes from one of the Higgs VEVs and the other is present 
in the Yukawa coupling matrices.  The quasimanifest and pseudomanifest models may be recovered as 
limits of our model by setting one or the other of these two fundamental phases
to zero or $\pi$.

The remainder of the paper is structured as follows.  In Secs.~\ref{sec:model} and \ref{sec:numsoln}
we outline our model and method of solution.  
Sections~\ref{sec:LevI}, \ref{sec:LevII} and \ref{sec:LevIINum} discuss 
the various experimental constraints and show how these limit
the parameter space of the model.  Sec.~\ref{sec:concl} contains some concluding
remarks.  In Appendix~\ref{sec:montecarlo} we describe some details of our Monte Carlo algorithm.
Appendix~\ref{sec:appendixb} contains a discussion of the relative
sizes of the left- and right-handed CKM rotation angles as well as approximate expressions for 
the CKM phases $\delta_L$ and $\delta_R$ in terms of the fundamental parameters in the model.

\section{The Model}
\label{sec:model}

In this section we explain our notation and summarize some of the important formal aspects of our model.
In particular, Sec.~\ref{sec:tilrm} contains one of the central
results of this paper, namely that many of the apparent degrees of freedom in the
nonmanifest model may be eliminated by a suitably chosen unitary rotation.
This insight leads to a considerable simplification of our task and allows
us to proceed with our numerical work.  For the purpose of the work to follow,
we are mainly concerned with the Yukawa couplings between the quark and Higgs fields.
The Higgs sector contains many intriguing features, including
a Flavour Changing Neutral Higgs and a doubly-charged Higgs.  Minimal versions of
the model include a bidoublet Higgs field and a pair of either
doublet or triplet Higgs fields.  
The triplet fields tend to be favoured in the
literature, since they can lead quite naturally to very light neutrino masses through
the See-Saw Mechanism.

\subsection{Quark Mass Matrices in the Left-Right Symmetric Model}
\label{sec:qmm}

Let us begin by deriving the relationship between the quark mass matrices and
the left- and right-handed CKM matrices.  In order to do so, we must first
consider the Higgs fields.
The Left-Right Model is based on the gauge group
$SU(2)_L\times SU(2)_R\times U(1)$, with the symmetry being spontaneously broken down to $U(1)_\textrm{\scriptsize{em}}$
through the Higgs mechanism.  The left- and right-handed quark fields transform as
doublets under the unbroken gauge groups $SU(2)_L$ and $SU(2)_R$, respectively.  The
particular Left-Right Model that we consider contains a bidoublet Higgs field $\Phi\sim(2,\overline{2},0)$
as well as two triplet Higgs fields $\Delta_{L}\sim(3,1,2)$ and $\Delta_{R}\sim(1,3,2)$,
\begin{eqnarray}
	\Phi = \left(\begin{array}{cc}
		\phi_1^0 & \phi_2^+ \\
		\phi_1^- & \phi_2^0 \\
		\end{array}\right)\; ,~~
	\Delta_{L,R} = \left(\begin{array}{cc}
		\Delta^+_{L,R}/\sqrt{2} & \Delta^{++}_{L,R} \\
		\Delta^0_{L,R} & -\Delta^+_{L,R}/\sqrt{2} \\
		\end{array}\right) \; .
\end{eqnarray}
The bidoublet field couples to the quarks and leptons and is responsible for giving
them masses, while the right-handed triplet field is used to break the left-right 
symmetry at some high energy scale.  The VEVs for these 
fields may be parameterized as follows~\footnote{Some authors use ``$\delta_{L,R}$''
for the VEVs of $\Delta_{L,R}$, but we prefer to reserve the symbols $\delta_{L,R}$ for phases appearing
in the left- and right-handed CKM matrices.},
\begin{eqnarray}
	\langle\Phi\rangle = \left(\begin{array}{cc}
		\kappa & 0 \\
		0 & \kappa^\prime \\
		\end{array}\right) \; ,~~
	\langle \Delta_{L,R} \rangle = \left(\begin{array}{cc}
		0 & 0 \\
		v_{L,R} & 0 \\
		\end{array}\right) \; .
\end{eqnarray}
In order to reproduce observed electroweak phenomenology, one typically assumes that
$|v_R|\gg |\kappa|,|\kappa^\prime|\gg|v_L|$, in which case
$\kappa$ and $\kappa^\prime$ satisfy the constraint~\cite{langacker}
\begin{eqnarray}
	|\kappa|^2 + |\kappa^\prime|^2 \simeq \frac{2m_W^2}{g^2}\simeq(174.1~\textrm{GeV})^2.
		\label{eq:kkpr1}
\end{eqnarray}
Although $\kappa$ and $\kappa^\prime$
are both in principle complex, the only physically observable phase
comes from their product, $\kappa \kappa^\prime$.
(One can always use a gauge rotation to eliminate the second phase.)
We shall, for simplicity, take $\kappa$ to be real and positive, so that the observable
phase is carried by $\kappa^\prime$; i.e.,
\begin{eqnarray}
	\alpha_{\kappa^\prime} = \arg (\kappa^\prime) \; .
		\label{eq:alphadef}
\end{eqnarray}

The Yukawa couplings of the quarks to the bidoublet Higgs fields may be written in terms
of two $3\times3$ Hermitian matrices $F$ and $G$ as follows,
\begin{eqnarray}
	-{\mathcal L}_\textrm{\scriptsize Yukawa} = \overline{\psi}_{iL}^\prime\left(
		F_{ij} \Phi + G_{ij} \widetilde{\Phi}\right)\psi_{jR}^\prime 
		+ \textrm{h.c.} \; ,
	\label{eq:yuk}
\end{eqnarray}
where $\widetilde{\Phi}= \tau_2 \Phi^* \tau_2$ and where
the gauge eigenstates $\psi^\prime_{iL,R}$ are given by
\begin{eqnarray}
	\psi^\prime_{iL,R} = \left(\begin{array}{c}
		u^\prime_{iL,R}\\
		d^\prime_{iL,R}\\
		\end{array}\right) \;.
\end{eqnarray}
The Hermiticity of $F$ and $G$ helps ensure the left-right symmetry of the 
Lagrangian~\footnote{See, for example, the discussion in Ref.~\cite{langacker}.}.
Insertion of the bidoublet Higgs
VEVs into Eq.~(\ref{eq:yuk}) yields the up- and down-type
quark mass matrices
\begin{eqnarray}
	{\mathcal M}_u & = & \kappa F + \kappa^{\prime *} G \label{eq:massup}\\
	{\mathcal M}_d & = & \kappa^\prime F + \kappa^* G \label{eq:massdown}\; .
\end{eqnarray}
${\mathcal M}_u$ and ${\mathcal M}_d$ are, in principle, complex matrices, and
may be diagonalized by biunitary transformations, yielding
\begin{eqnarray}
	{\mathcal M}_u^\textrm{\scriptsize diag} & = & 
		V_L^{U\dagger}{\mathcal M}_uV_R^U \\
	{\mathcal M}_d^\textrm{\scriptsize diag} & = & 
		V_L^{D\dagger}{\mathcal M}_dV_R^D,
\end{eqnarray}
where ${\mathcal M}_u^\textrm{\scriptsize diag} \equiv \textrm{diag}(m_u,m_c,m_t)$
and ${\mathcal M}_d^\textrm{\scriptsize diag} \equiv \textrm{diag}(m_d,m_s,m_b)$.
One can always choose the unitary rotation matrices $V_{L,R}^{U,D}$ in
such a way that the elements of the diagonalized mass matrices are real and positive.

With the diagonalization matrices in hand, the charged-current
Lagrangian may be written in terms of the (unprimed) quark mass eigenstates,
\begin{eqnarray}
	{\mathcal L}_{CC} & = &
		-\frac{g}{\sqrt{2}}\overline{u}_{L}V_{L}^{CKM}\gamma_\mu d_L W_L^{\mu+} 
		-\frac{g}{\sqrt{2}}\overline{u}_{R}V_{R}^{CKM}\gamma_\mu d_R W_R^{\mu+}
		+ \textrm{h.c.},
\end{eqnarray}
where the generation indices have been suppressed and where
we have taken the left- and right-handed weak coupling constants
to be equal, $g_L=g_R \equiv g$.  The
left- and right-handed CKM matrices in the above expression are given by
\begin{eqnarray}
	V_{L}^{CKM} & = & B^\dagger V_L^{U\dagger}V_L^D\widetilde{B} 
		\label{eq:VLCKM}\\
	V_{R}^{CKM} & = & B^\dagger V_R^{U\dagger}V_R^D\widetilde{B}.
		\label{eq:VRCKM}
\end{eqnarray}
The matrices $B$ and $\widetilde{B}$ are diagonal phase matrices that are used to
rotate as many phases as possible out of the left-handed CKM matrix (and hence into the
right-handed CKM matrix), leaving $V_{L}^{CKM}$ in its ``standard'' form with only one
CP-violating phase $\delta_L$~\cite{pdg2000},
\begin{eqnarray}
	V_{L}^{CKM}(\theta_{12},\theta_{23},\theta_{13},\delta_L) & & \nonumber \\
		&\!\!\!\!\!\!\!\! = & \left(\begin{array}{ccc}
		c_{12}c_{13} & s_{12}c_{13} & s_{13}e^{-i\delta_L} \\
		-s_{12}c_{23}-c_{12}s_{23}s_{13}e^{i\delta_L} &
			c_{12}c_{23}-s_{12}s_{23}s_{13}e^{i\delta_L} &
			s_{23}c_{13} \\
		s_{12}s_{23}-c_{12}c_{23}s_{13}e^{i\delta_L} &
			-c_{12}s_{23}-s_{12}c_{23}s_{13}e^{i\delta_L} &
			c_{23}c_{13} \\
		\end{array}\right) \! .
	\label{eq:vlckm}
\end{eqnarray}
In the above expression, $s_{ij} \equiv \sin\theta_{ij}$, and all sines and cosines
are taken to be non-negative.
The left-handed phase $\delta_L$ is the usual CKM phase and is the sole source of CP violation
within the context of the SM.  It is also very nearly equal to the
(perhaps more familiar) angle $\gamma$,
\begin{eqnarray}
	\delta_L \simeq \gamma = \arg\left(-\frac{V_{L_{ud}}^{CKM} V_{L_{ub}}^{CKM*}}
			{V_{L_{cd}}^{CKM} V_{L_{cb}}^{CKM*}} \right) \; .
\end{eqnarray}
With the above parametrization for $V_L^{CKM}$, $V_R^{CKM}$ has six non-removable
phases.  A convenient parameterization for $V_R^{CKM}$ is as follows,
\begin{eqnarray}
	V_{R}^{CKM}= K \; V_L^{CKM}(\theta_{12}^R,\theta_{23}^R,\theta_{13}^R,\delta_R)
			\; \widetilde{K}^\dagger \; ,
\end{eqnarray}
where the right-handed rotation angles $\theta_{ij}^R$ are again taken to be
in the first quadrant, so that all sines and cosines are non-negative.
The diagonal matrices $K$ and $\widetilde{K}$ contain five of the six non-removable
phases in $V_R^{CKM}$,
\begin{eqnarray}
	K & = & \textrm{diag}(e^{i\rho_1},e^{i\rho_2},e^{i\rho_3}) \\
	\widetilde{K} & = & \textrm{diag}(1,e^{i\eta_2}, e^{i\eta_3}),
		\label{eq:ktilde}
\end{eqnarray}
with the sixth phase being $\delta_R$, the right-handed analogue of $\delta_L$.
It is important to note that the seven non-removable phases distributed among
$V_L^{CKM}$ and $V_R^{CKM}$ are not, in general, all independent.  
In our model, for example, the seven phases are functions of only two ``basic'' phases ($\alpha_{\kappa^\prime}$
and $\beta_{23}$, see Eqs.~(\ref{eq:alphadef}) and (\ref{eq:FG})).

There are several ways to achieve CP violation within the Left-Right Model, and
it is useful to enumerate these.
\begin{enumerate}
\item \textit{(Quasi)manifest left-right symmetry.} The simplest
case occurs when CP is broken explicitly by the Yukawa
couplings of the quarks to the Higgs fields.  The product $\kappa \kappa^\prime$ is real 
(so that $\alpha_{\kappa^\prime} = 0$ or $\pi$), but $F$ and $G$ (and hence
the mass matrices themselves) are complex and Hermitian.  
In this case one has $V_{Rij}^{CKM} = \pm V_{Lij}^{CKM}$, a situation referred to as
``quasimanifest left-right symmetry'' in Ref.~\onlinecite{deshpande}.  ``Manifest left-right
symmetry'' refers to the special case in which the ``$+$'' sign occurs for each of the nine
elements of the matrices.
\item \textit{Pseudomanifest left-right symmetry.} 
One can also allow the product $\kappa \kappa^\prime$ to carry the CP-violating phase and require
$F$ and $G$ to remain real (and symmetric).  In this case CP is broken spontaneously
and the resulting mass matrices are complex symmetric.  The
left- and right-handed CKM matrices in this case satisfy the relation 
$V_{R}^{CKM} = A V_{L}^{CKM*}\widetilde{A}^\dagger$, where $A$ and $\widetilde{A}$ are
diagonal phase matrices; i.e., elements of the two matrices are equal in magnitude, but
could have different phases.  This case is often referred to as ``pseudomanifest left-right symmetry.''
\item \textit{Nonmanifest left-right symmetry.}
In the most general case (considered in the present work),
one allows both the product $\kappa \kappa^\prime$ and the matrices
$F$ and $G$ to be complex, while maintaining the Hermiticity of $F$ and $G$.
In this case, the mass matrices are in principle arbitrary complex matrices and
the left- and right-handed CKM matrices have no special relations to each other.
In particular, unlike in the previous cases, the rotation angles in $V_{L}^{CKM}$
and $V_{R}^{CKM}$ need not be equal.  Many authors have considered the general case,
and have made various ansatzes for the form of $V_R^{CKM}$.  Langacker and Sankar, 
for example, argued that a relatively light right-handed $W$ could be accommodated
if $V_R^{CKM}$ took on one of the forms~\cite{langacker}
\begin{eqnarray}
	V_{R(A)}^{CKM} = \left(\begin{array}{ccc}
		1 & 0 & 0 \\
		0 & c & \pm s \\
		0 & s & \mp c \\
		\end{array}\right)\; ,~~~~~
	V_{R(B)}^{CKM} = \left(\begin{array}{ccc}
		0 & 1 & 0 \\
		c & 0 & \pm s \\
		s & 0 & \mp c \\
		\end{array}\right)\; .
	\label{eq:vrckmlangacker}
\end{eqnarray}
The mixing angles in these expressions are clearly quite different from those
of the left-handed CKM matrix.
\end{enumerate}

In the present work we use the experimentally determined quark masses and left-handed 
rotation angles to delineate the various possibilities for the right-handed CKM matrix.
We find that the right-handed rotation
angles are very similar in magnitude to their left-handed counterparts
(see Fig.~\ref{fig:rotnangles} and Appendix~\ref{sec:appendixb}), ruling
out the forms given in Eqs.~(\ref{eq:vrckmlangacker}), at least within the context of the class
of models considered in this paper.

\subsection{A Top-Inspired Left-Right Model}
\label{sec:tilrm}

In the present work we consider a particular version of the Left-Right Model
that is inspired by the relatively large masses
of the third generation quarks as well as by the small mixing that
exists between the third generation and the first two in $V_L^{CKM}$.  It was pointed
out many years ago that small 1-3 and 2-3 mixings follow naturally if one
takes the ratio $\kappa^\prime/\kappa$ to be of order
$m_b/m_t$~\cite{ecker80,ecker81,ecker85,frere}.  This scenario was considered
recently in the ``Spontaneously Broken Left-Right Model'' (SB-LR),
where the authors chose to fix the ratio at $m_b/m_t$~\cite{ball1,ball2},
\begin{eqnarray}
	\left|\frac{\kappa^\prime}{\kappa}\right|=\frac{m_b}{m_t}\; .
	\label{eq:kkpr2}
\end{eqnarray}
The SB-LR is an example of a pseudomanifest
left-right symmetric model.  It contains many attractive features, such as spontaneous CP violation
(arising from a single CP-odd phase)
and upper limits on the Higgs and right-handed $W$ masses.  
However, according to Ref.~\onlinecite{ball1}, the SB-LR
predicts $\sin 2\beta_{CKM}^\textrm{\scriptsize eff}\alt 0.1$.
This value is difficult to reconcile with recent
precision measurements, which give
$\sin 2\beta_{CKM}^\textrm{\scriptsize eff}= 0.79 \pm 0.11$~\cite{babar2001,belle2001,cdf2000}.
\footnote{In determining the weighted average and its uncertainty, we use only the BABAR and BELLE
results.  The earlier CDF and LEP results are in agreement with
the value we use.  Inclusion of the CDF and LEP results would have only a very slight effect
on the weighted average and its uncertainty.}
The SB-LR, although quite attractive, is somewhat tightly constrained because it does not allow for 
explicit CP violation in the Yukawa couplings.  In the present work we
retain the constraint given in Eq.~(\ref{eq:kkpr2}), but generalize 
the mode of CP violation by allowing the Yukawa coupling matrices $F$ and
$G$ to be (in principle) arbitrary $3\times 3$ Hermitian matrices.
At first glance it might appear that our generalization
would hopelessly complicate matters by adding many new CP-odd phases.
As we demonstrate below, however, it is possible to simplify the forms of $F$ and $G$ (without
any loss of generality) in such a way that our model contains only one new phase
compared to either the SB-LR or the SM (see Eq.~(\ref{eq:FG}) below).

It is useful to consider the number of degrees of freedom contained within
the quark mass matrices, as well as the experimental constraints
that may be placed on these.  
The most general expressions for the mass matrices in the Left-Right Model 
are given in Eqs.~(\ref{eq:massup}) and (\ref{eq:massdown}),
where $\kappa$ is real, $\kappa^\prime$ is complex and 
$F$ and $G$ are Hermitian. 
The magnitudes of $\kappa$ and $\kappa^\prime$ are
fixed within our model (see Eqs.~(\ref{eq:kkpr1}) and (\ref{eq:kkpr2})) and the phase
of $\kappa$ has been gauged away.  This leaves just one degree of freedom among
$\kappa$ and $\kappa^\prime$, namely the phase of $\kappa^\prime$.
Since $F$ and $G$ are both Hermitian $3\times 3$ matrices, it would appear
at first glance that there are a total of 18 degrees of freedom contained
in $F$ and $G$, for a total of 19 degrees of freedom within the quark mass matrices.
It should be noted that six of these degrees of freedom would be new phases
when compared to the SB-LR.  
On the experimental side, there are nine direct constraints on the quark mass
matrices (the so-called ``Level I'' constraints below).  These come from the six
quark masses and the three left-handed rotation angles.
The mass matrices would thus appear to be severely underconstrained, a situation that would
not be ideal for a numerical study of the model.
Fortunately, however, eight of the apparent degrees of freedom within $F$ and $G$ --
including five of the six phases -- may be rotated away by performing a unitary rotation in flavour space.
This leaves 11 degrees of freedom with nine direct constraints placed upon them, a much more
agreeable situation from the standpoint of solving the model numerically.

A rotation of all up, down, left and right quark fields by the same unitary 
matrix is unobservable, so one may perform the rotations $\mathcal{U} F \mathcal{U}^\dagger$
and $\mathcal{U} G \mathcal{U}^\dagger$ with no observable consequences.  In particular, we
can use $\mathcal U$ to diagonalize $F$, which yields all real eigenvalues (since $F$ is Hermitian).
It is also possible to change the overall signs of both $F$ and $G$ (simultaneously)
with no observable consequences, so that a given element of the diagonalized version of $F$ may always
be taken to be positive.  Having diagonalized $F$,
we may perform a diagonal phase rotation (which does not affect $F$) in order
to eliminate two phases in $G$.  
As a result, we may quite generally take $F$ and $G$ to be of the form
\begin{eqnarray}
	F = \left(\begin{array}{ccc}
		f_{11} & 0 & 0 \\
		0 & f_{22} & 0 \\
		0 & 0 & f_{33} \\
		\end{array}\right) \; ,~~~~
	G = \left(\begin{array}{ccc}
		g_{11} & g_{12} & g_{13} \\
		g_{12} & g_{22} \;\; & g_{23} e^{i\beta_{23}} \\
		g_{13} \;\; & g_{23} e^{-i\beta_{23}} & g_{33} \\
		\end{array}\right) \; ,
	\label{eq:FG}
\end{eqnarray}
where all $f_{ii}$ and $g_{ij}$ are real, $f_{33}\geq 0$ and $g_{ij}\geq 0$ for $i\neq j$,
and where the sole non-trivial
phase, $\beta_{23}$, has (arbitrarily) been placed in the 2-3 element of $G$~\footnote{Note that 
if any of the off-diagonal elements of $G$ are zero then the phase $\beta_{23}$ may also be
rotated away, yielding once again the pseudomanifest case.}.  One may also assume without loss of
generality that $f_{33}\geq |f_{22}| \geq |f_{11}|$.
It should be emphasized that the forms given for $F$ and $G$ in Eq.~(\ref{eq:FG}) are completely
general: the mass matrices of any Left-Right Model (containing only a 
single bidoublet Higgs field) may be written in terms of only two
CP-odd phases, which may be parameterized as $\alpha_{\kappa^\prime} = \arg(\kappa^\prime)$ and
$\beta_{23} = \arg(G_{23})$~\footnote{The results of the preceding few paragraphs are
easily generalized to the case of $N$ quark generations.  In that case, if one wishes
to put as few phases as possible in the left-handed CKM matrix, that number would be
$(N-1)(N-2)/2$, leaving $N(N+1)/2$ non-removable phases for the right-handed CKM
matrix.  Regarding ``fundamental'' or ``basic'' phases in the model, there
would be 1 non-removable phase among the bidoublet Higgs VEVs and 
$(N-1)(N-2)/2$ non-removable phases in the Yukawa matrices.  These latter
phases could all be placed in ``$G$,'' as has been done in the present work.}.

Before proceeding to the numerical solution of our model,
let us first obtain order-of-magnitude estimates for the elements of the Yukawa coupling
matrices $F$ and $G$.  Our task is greatly simplified by the constraints 
in Eqs.~(\ref{eq:kkpr1}) and (\ref{eq:kkpr2}).  These imply, in an interesting coincidence, that
\begin{eqnarray}
	\kappa & \sim & \sqrt{\frac{2m_W^2}{g^2}} \simeq 174.1~\textrm{GeV} \sim m_t,\\
	|\kappa^\prime| & \sim & \frac{m_b}{m_t}\sqrt{\frac{2m_W^2}{g^2}} \sim m_b \; .
	\label{eq:kappaprime}
\end{eqnarray}
Consider first the expression given for ${\mathcal M}_u$ in Eq.~(\ref{eq:massup}).
Since $\kappa\gg |\kappa^\prime|$, we have, to a first approximation, that 
${\mathcal M}_u\sim \kappa F$.  (The elements of $G$ do contribute somewhat to 
${\mathcal M}_u$, but such contributions are suppressed by the small size of $\kappa^\prime$.)
Thus, to a first approximation one might expect
\begin{eqnarray}
	F \sim \left(\begin{array}{ccc}
		\mathcal{O}\left(\frac{m_u}{m_t}\right) & 0 & 0 \\
		0 & \mathcal{O}\left(\frac{m_c}{m_t}\right) & 0 \\
		0 & 0 & \mathcal{O}(1) \\
		\end{array}\right) \; .
	\label{eq:fapprox}
\end{eqnarray}
We will see below that the above hierarchy is indeed observed by the numerical solutions.
The down quark mass 
matrix, ${\mathcal M}_d=\kappa^\prime F + \kappa^* G$, 
yields insight both into the elements of $G$ and into 
the sizes of the left- and right-handed rotation angles.
Consider first the 3-3 element of ${\mathcal M}_d$.  From 
Eqs.~(\ref{eq:kappaprime}) and (\ref{eq:fapprox}), we see that $\kappa^\prime F_{33}$ is naturally
of order $m_b$.  If $\kappa^* G_{33}$ is also to be of order $m_b$,
then $G_{33}$ must be of order $m_b/m_t$.  The other two elements of $F$ give relatively small
contributions to their respective down-type quark masses (since $\kappa^\prime F_{ii}$
is ``too small'' in those two cases), so $\kappa^* G$ has primary responsibility for the
masses of the first and second generation down-type quarks.  This confirms our initial assertion
that $\kappa^{\prime *}G$ would give a relatively small contribution to 
${\mathcal M}_u$.  To the extent that ${\mathcal M}_u\sim \kappa F$ is a good approximation,
we have that  $V_L^U$ and $V_R^U$ are approximately diagonal matrices.
Thus $V_L^{CKM}$ and $V_R^{CKM}$ are
essentially determined by $V_L^D$ and $V_R^D$ (see Eqs.~(\ref{eq:VLCKM}) and
(\ref{eq:VRCKM})).
In a somewhat poorer approximation, we 
could also almost take ${\mathcal M}_d\sim \kappa^* G$, except 
for the 3-3 element.  Noting that in our notation $\kappa$ is real, we then have that
${\mathcal M}_d$ is ``almost'' Hermitian, except for its 3-3 element.  If ${\mathcal M}_u$
and ${\mathcal M}_d$ were exactly Hermitian, then
the left- and right-handed CKM matrices would have obeyed the ``quasimanifest'' condition
noted above, $V_{Rij}^{CKM} = \pm V_{Lij}^{CKM}$.  In practice, we find that
the left- and right-handed rotation angles do agree reasonably well (see Fig.~\ref{fig:rotnangles}, below),
that $\delta_R \approx \delta_L$ and that the right-handed phases 
corresponding to the first and second generations ($\rho_1$, $\rho_2$ and $\eta_2$)
are all close to zero or $\pi$.  The right-handed phases corresponding to the third generation
($\rho_3$ and $\eta_3$), by way of contrast, can take on any values.  (See Fig.~\ref{fig:LRphases} below.)
Appendix~\ref{sec:appendixb} contains some further
discussion along these lines, along
with approximate analytical relations for various CKM mixing angles and phases.

\section{Numerical Solution of the Model -- An Overview}
\label{sec:numsoln}

We turn now to a brief explanation of the numerical solution of our model.
As noted above, the 
quark mass matrices in our model are described by eleven real ``input'' parameters, namely
\begin{eqnarray}
	f_{ii},~ g_{ij},~ \alpha_{\kappa^\prime},~\textrm{and}~\beta_{23}~~~~~
			\textrm{(11 input parameters).} \label{eq:inputs}
\end{eqnarray}
The model as a whole contains two more parameters --  $M_2$ (the mass
of the mostly right-handed heavy $W$) and $M_H$ (the Higgs mass scale~\footnote{We adopt the
convention of several previous authors and assume that the various non-standard physical Higgs
bosons all have the same masses.}) -- bringing the total number of ``input'' parameters
in the model to thirteen.  The goal of the numerical work is to find combinations
of the various input parameters that yield acceptable values for quantities
that are known experimentally, such as the quark masses, left-handed
rotation angles and $\epsilon_K$.

Each possible set of input parameters may be used to form trial mass matrices,
which may then be diagonalized to yield the physical quark masses and the left-
and right-handed CKM matrices, as described above in Sec.~\ref{sec:qmm}.
The resulting quark masses and left-handed rotation angles (and possibly other
quantities, such as $\epsilon_K$, etc.) may then be compared to their
known experimental values.  A ``solution'' refers to a
set of input parameters that satisfies all relevant experimental constraints
to within some prescribed tolerance.  In principle one could search
the entire input parameter space for such solutions, but this is not possible in practice
since the parameter space contains many dimensions, each of which contains a continuum
of values.  Fortunately, we can narrow down the parameter space to several promising regions
by using the reasoning outlined in Sec.~\ref{sec:tilrm}.  But even with the parameter
space pared down in this manner, it would still be very difficult (and inefficient) to find solutions
by simply slicing up the multi-dimensional space into many small hypercubes.  To overcome this problem
we have devised an adaptive Monte Carlo routine (described in detail in Appendix~\ref{sec:montecarlo})
that is able to zoom in on solutions with relatively high efficiency.

\begin{table*}
\caption{Values used in the determination of Level I and II constraints.  The first twelve
rows correspond to the Level I constraints, while the addition of the last five
rows yields the Level II constraints.  The last two constraints are 
not included in the calculation of $\chi_{(II)}^2$, but are imposed
as ``cuts'' after $\chi_{(II)}^2$ has satisfied the required tolerance.  Quark masses
are in GeV and are evaluated at the energy scale $m_Z$~\cite{koide};
$\Delta m_{B_d}$ is in ps$^{-1}$.  Constraints 7-9 are
taken from Ref.~\cite{ball1}.  The central values,
uncertainties and limits listed in this table
are discussed in detail in Secs.~\ref{sec:LevI} and \ref{sec:LevII}.}
\label{tab:chi2}
\begin{ruledtabular}
\begin{tabular}{cccc}
	$i$ & $y_i$ & central value ($y_i^\textrm{\scriptsize{exp}}$) 
		& uncertainty ($\sigma_i$) or limits \\ \colrule
	1 & $m_u$ & $2.33\times10^{-3}$ & $0.45\times 10^{-3}$ \\
	2 & $m_d$ & $4.69\times10^{-3}$ & $0.66\times 10^{-3}$ \\
	3 & $m_c$ & 0.685 & 0.061 \\
	4 & $m_s$ & 0.0934 & 0.0130 \\
	5 & $m_t$ & 181 & 13 \\
	6 & $m_b$ & 3.00 & 0.11 \\
	7 & $\sin\theta_{12}$ & 0.2200  & 0.0030 \\
	8 & $\sin\theta_{23}$ & 0.0395  & 0.0017 \\
	9 & $\sin\theta_{13}$ & 0.0032  & 0.0008 \\
	10 & $m_u/m_d$ & 0.497  & 0.119 \\
	11 & $m_s/m_d$ & 19.9  & 3.9 \\
	12 & $(m_s-(m_u+m_d)/2)/(m_d-m_u)$ & 38.1  & 14.1 \\ \colrule
	13 & $\epsilon_K$ & $2.28\times 10^{-3}$  & $[0.46+3.5 \times
		(1.0~\textrm{TeV}/M_2)^2]\times 10^{-3}$ \\
	14 & $\sin 2\beta_{CKM}^\textrm{\scriptsize eff}$ & 0.79 & 0.11 \\
	15 & $\Delta m_{B_d}$ 	& 0.472 & 0.190 \\
	-- & $\Delta m_{B_s}/\Delta m_{B_d}$ & -- & $\Delta m_{B_s}/\Delta m_{B_d}\ge 27.2$ \\
	-- & $\Delta m_{K}$ & -- &  
		$-1\le 2\; \textrm{Re}(M_{12}^{LR})/\Delta m_K^\textrm{\scriptsize{exp}}\le 1$ \\
	   & & & \textrm{and~~}$\textrm{Re}(M_{12})> 0$ \\
\end{tabular}
\end{ruledtabular}
\end{table*}

When discussing constraints satisfied by ``solutions''
in our model, it is useful to distinguish between minimal and higher-level constraints.
Table~\ref{tab:chi2} lists the various constraints employed.
The twelve ``Level I'' constraints take into account the quark masses and left-handed
rotation angles, while the ``Level II'' constraints include additional
experimental inputs from the neutral kaon and $B$ systems.
Sections~\ref{sec:LevI} and \ref{sec:LevII} discuss these experimental constraints
in detail.
When the Level I (or II) constraints
have been satisfied to within the required tolerance, we call the set of
input parameters a Level I (or II) solution.  In order to judge how close
a given set of input parameters is to being a solution, we define
a $\chi^2$ for each of the constraints, and then sum these up to obtain
a $\chi^2$ for the appropriate constraint level; i.e.,
\begin{eqnarray}
	\chi_i^2 & = & \frac{(y_i-y_i^\textrm{\scriptsize{exp}})^2}{\sigma_i^2} \\
	\chi_{(I)}^2 & = & \sum_{i=1}^{12} \chi_i^2 \label{eq:conLevI}\\
	\chi_{(II)}^2 & = & \chi_{(I)}^2 + \sum_{i=13}^{15} \chi_i^2
		\label{eq:conLevII}
\end{eqnarray}
In these expressions $y_i^\textrm{\scriptsize{exp}}$ represents the known ``experimental''
value for the $i^\textrm{\scriptsize{th}}$ constraint, $y_i$ represents
the value produced theoretically using the given set of input parameters
and $\sigma_i$ represents the uncertainty (experimental and/or theoretical) for the constraint.
A set of input parameters is termed a solution if each of the relevant
$\chi_i^2$ is less than or equal to one.  In the case of a Level II solution,
a few additional ``cuts'' must also be passed, as will be described below.
Appendix~\ref{sec:montecarlo} details the method by which our adaptive Monte Carlo
routine attempts to minimize $\chi_{(I)}^2$ or $\chi_{(II)}^2$ 
as it searches the input parameter space for solutions.

\begin{table*}
\caption{Ranges used for the input parameters.  The various parameters are defined in 
Eqs.~(\ref{eq:alphadef}) and (\ref{eq:FG}).  The Monte Carlo
procedure that searches for solutions takes the initial range for the $i^\textrm{\scriptsize{th}}$
parameter to be $x_{i,0}^\textrm{\scriptsize cent} \pm \Delta_i$.  The reader is referred to Appendix~\ref{sec:montecarlo} for
more details on the adaptive Monte Carlo algorithm.}
\label{tab:inputranges}
\begin{ruledtabular}
\begin{tabular}{cccc}
	$i$ & $x_i$ & initial central value ($x_{i,0}^\textrm{\scriptsize{cent}}$) 
		& initial range ($\Delta_i$) \\ \colrule
	1 & $f_{11}$ & 0 & $10^{-4}$ \\
	2 & $f_{22}$ & 0 & $0.03$ \\
	3 & $f_{33}$ & 1.04 & $0.25$ \\
	4 & $g_{11}$ & 0 & $3\times10^{-4}$ \\
	5 & $g_{22}$ & 0 & $3\times10^{-3}$ \\
	6 & $g_{33}$ & 0 & $0.06$ \\
	7 & $g_{12}$ & $10^{-3}$ & $10^{-3}$ \\
	8 & $g_{13}$ & $10^{-3}$ & $10^{-3}$ \\
	9 & $g_{23}$ & $0.01$ & $0.01$\\ \colrule
	10 & $\alpha_{\kappa^\prime}$ & $\pi$  & $\pi$ \\
	11 & $\beta_{23}$ & $\pi$ & $\pi$ \\
\end{tabular}
\end{ruledtabular}
\end{table*}

Table~\ref{tab:inputranges} summarizes the ranges of the input parameters
that yield viable values for the quark masses and left-handed
CKM rotation angles.
Denoting the various input parameters by $x_i$ ($i=1,2,\ldots, 11$), we have found that solutions
for each of the $x_i$ lie within the range
\begin{eqnarray}
	x_{i,0}^\textrm{\scriptsize cent} \pm \Delta_i \; ,
\end{eqnarray}
where $x_{i,0}^\textrm{\scriptsize cent}$ and $\Delta_i$ 
are given in the table.  The ranges listed in the table
correspond to the initial regions that the adaptive Monte Carlo
algorithm uses as it searches for solutions.
These ranges are consistent with the arguments
made above in Sec.~\ref{sec:tilrm}.
(Note that the Level I or II solutions themselves form a subspace of the regions indicated in the table.)

In the following three sections we discuss the two levels of constraint in detail
and show the numerical results in each case.  We also provide comparisons, where
appropriate, with work performed by previous authors.

\section{Level I Constraints: Quark Masses and Rotation Angles}
\label{sec:LevI}

It is useful to begin by applying only the Level I constraints, that is, only those constraints that
involve the quark masses and left-handed CKM rotation angles in Table~\ref{tab:chi2}.
This analysis will help in the comparison to work performed by previous authors and will help
clarify the mathematical structure of the model.  
Note that any Level I 
solution may in principle be combined with values for $M_2$ and $M_H$ (the masses
of the predominantly right-handed gauge boson and the Higgs boson, respectively)
and checked to see
if the combination satisfies the Level II constraints.  This is not a particularly efficient
method of finding Level II solutions if $M_2$ and $M_H$ are fixed at certain values,
but does work reasonably well if many different pairs of masses are used.

Let us first discuss the Level I constraints themselves.  
The quark masses and uncertainties listed in the table are evaluated at the scale 
$m_Z$~\cite{koide}.  These values were also used to determine the quark mass ratios 
and their uncertainties.  The resulting
ranges for the mass ratios are reasonably consistent with those quoted in 
Ref.~\onlinecite{pdg2000}, except that our range for the 
third ratio is somewhat larger than that quoted in Ref.~\onlinecite{pdg2000}.  
The central values and uncertainties of
the sines of the three rotation angles are taken from Ref.~\onlinecite{ball1}.

\begin{figure}
\resizebox{\textwidth}{!}{\includegraphics{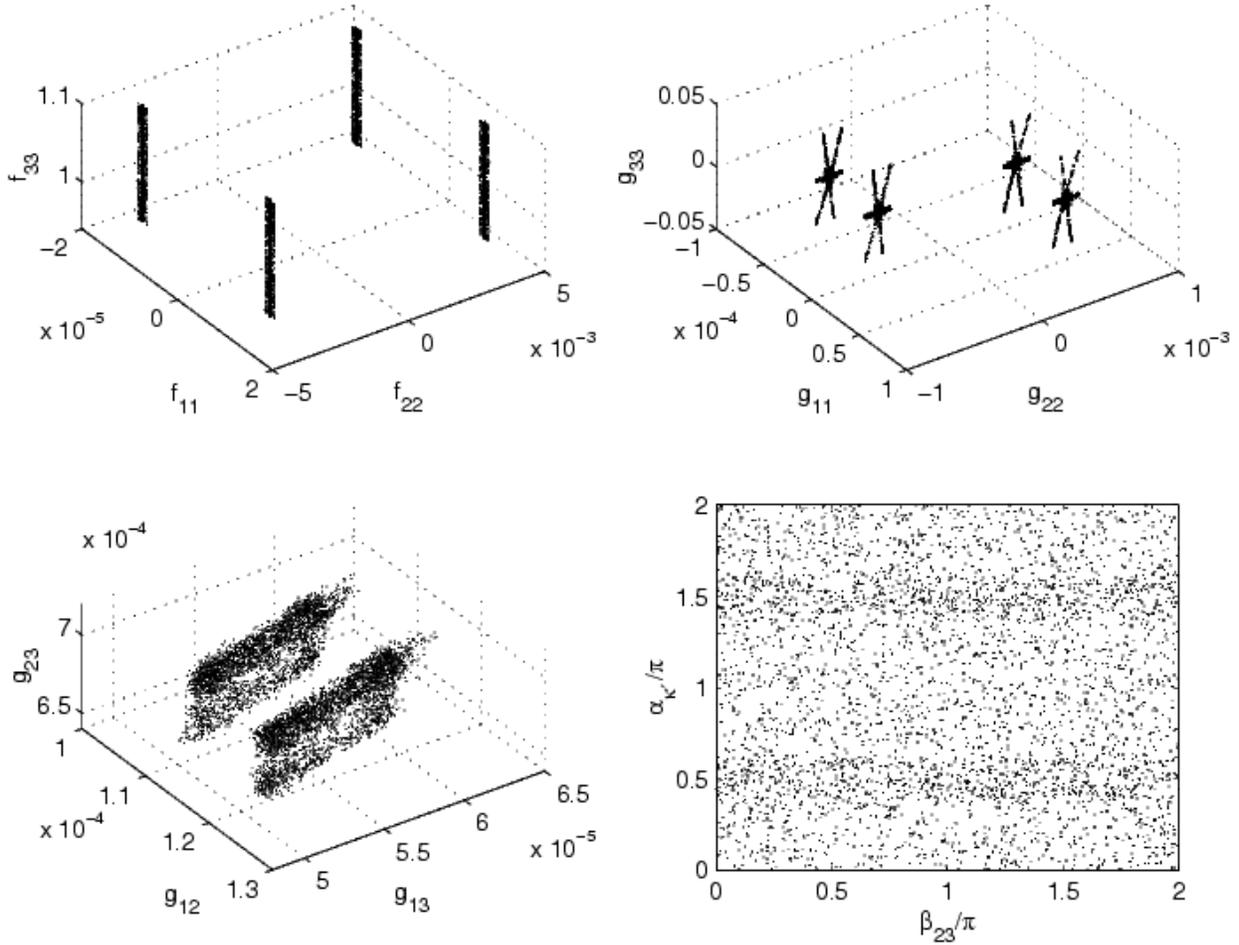}}
\caption{Regions of the input parameter space yielding viable quark masses and left-handed mixing angles.
For each ($\beta_{23}$, $\alpha_{\kappa^\prime}$) pair,
the adaptive Monte Carlo procedure was used to find
a set of values for the nine parameters $f_{ii}$ and $g_{ij}$ that satisfied the
Level I constraints.  No extra constraints from the $K$ or $B$ systems have been imposed.
The reader is referred to Appendix~\ref{sec:montecarlo} for details concerning the numerical procedure
used to find solutions.}
\label{fig:FGspace}
\end{figure}
\begin{figure}
\resizebox{\textwidth}{!}{\includegraphics{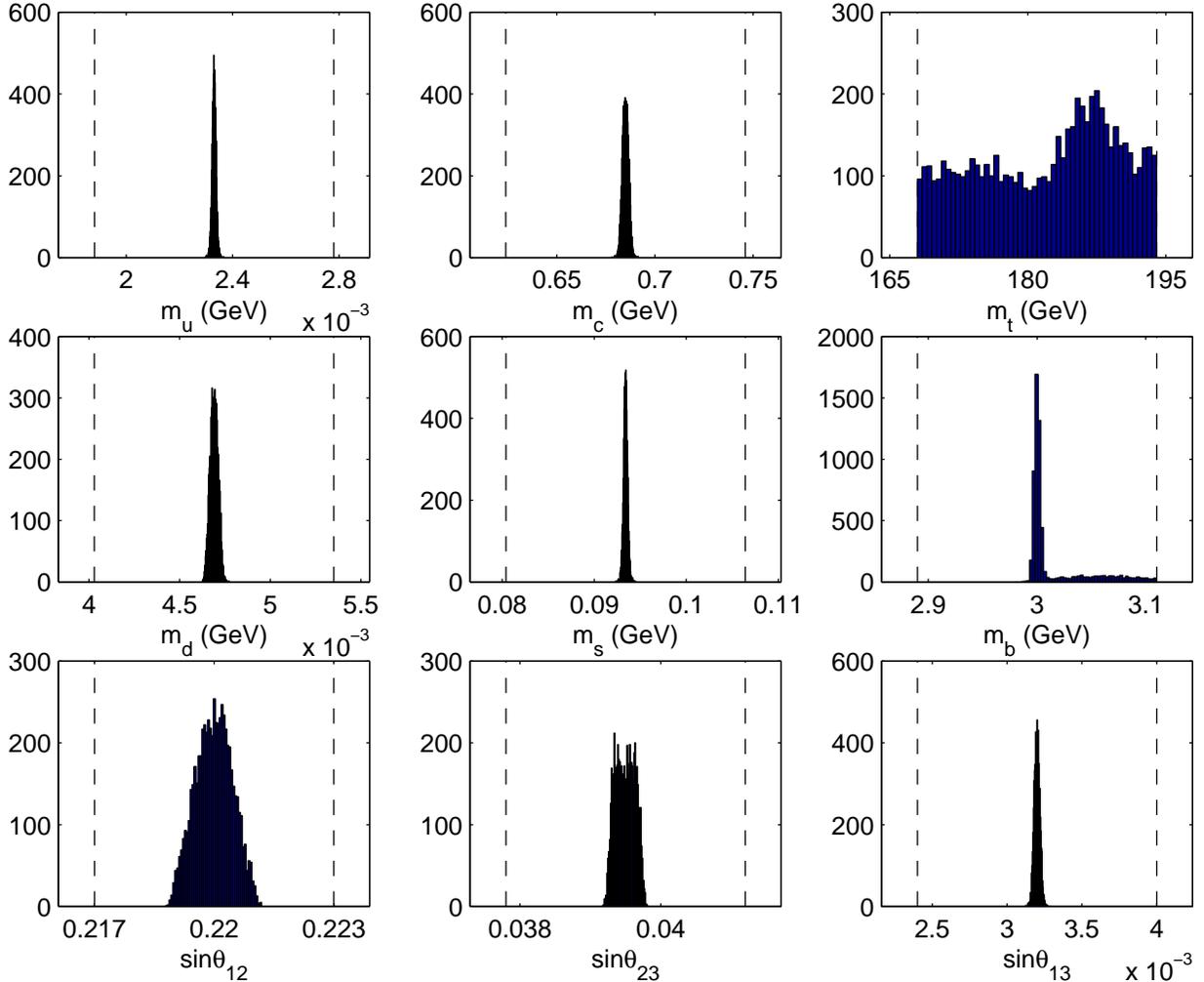}}
\caption{Frequency spectrum of quark masses and rotation angles for the data set
plotted in Fig.~\ref{fig:FGspace}.  The dashed vertical lines
indicate the values $y_i^\textrm{\scriptsize exp}\pm \sigma_i$.
The vertical axis gives the number of observations in each bin.}
\label{fig:masshist}
\end{figure}
\begin{figure}
\resizebox{\textwidth}{!}{\includegraphics{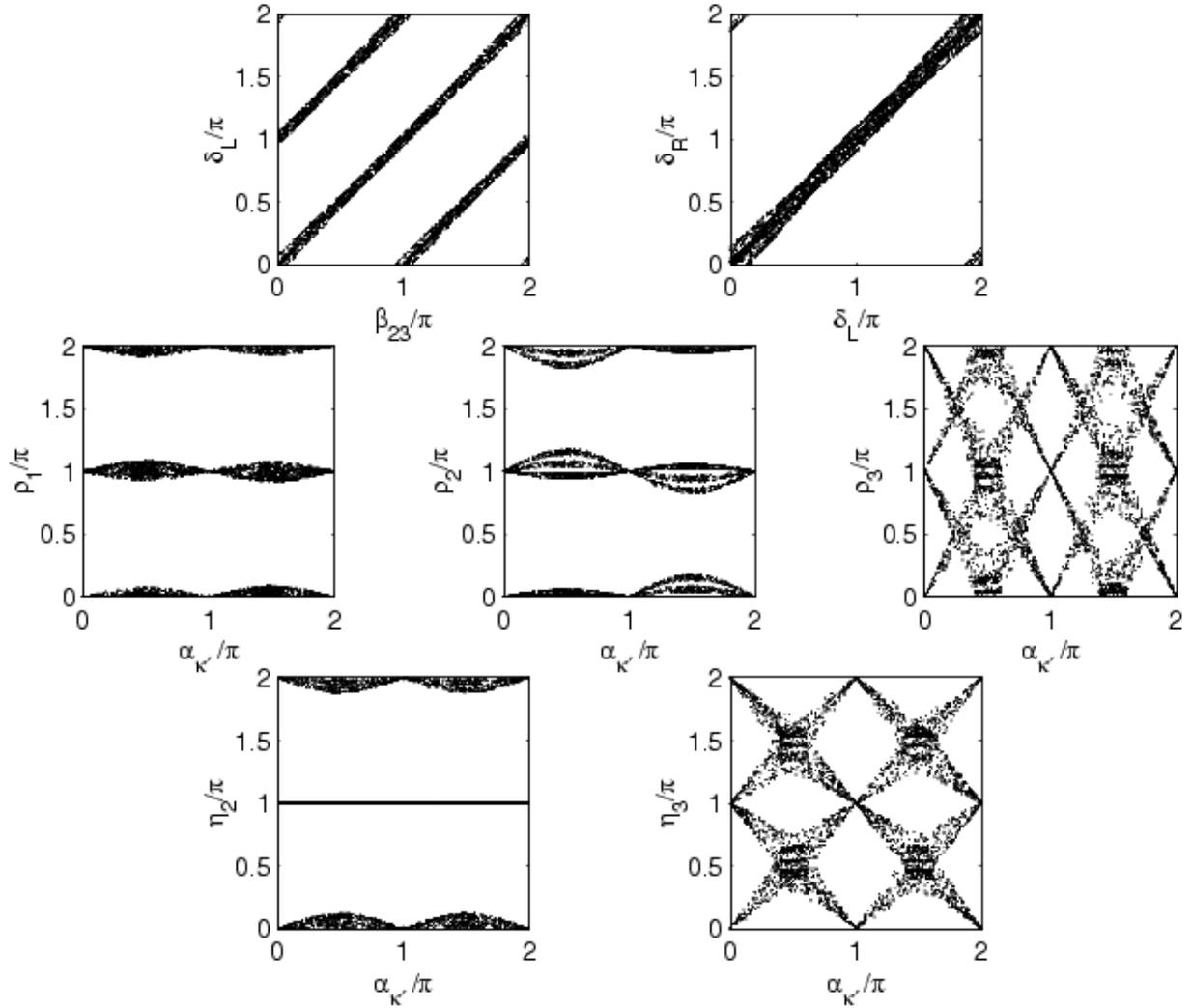}}
\caption{Left- and right-handed CKM phases for the set of input parameters shown
in Fig.~\ref{fig:FGspace}.  Note that the horizontal axes vary from plot
to plot.  Also, recall the definitions of the phases $\alpha_{\kappa^\prime}$
and $\beta_{23}$: $\alpha_{\kappa^\prime}=\arg(\kappa^\prime)$ and 
$\beta_{23}=\arg(G_{23})$.}
\label{fig:LRphases}
\end{figure}
\begin{figure}
\resizebox{\textwidth}{!}{\includegraphics{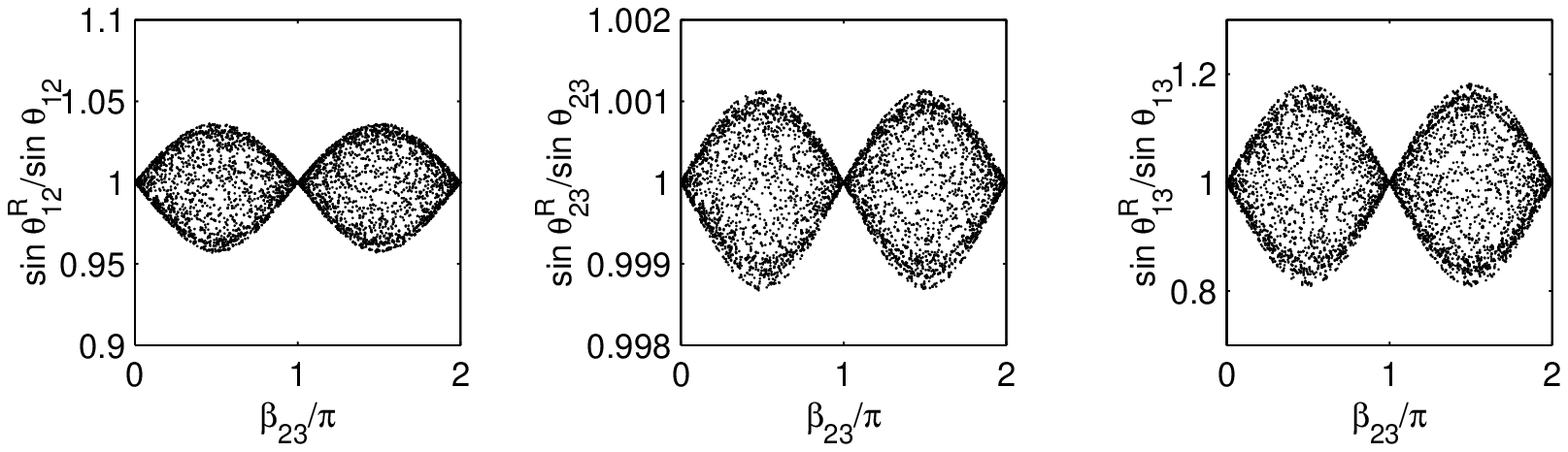}}
\caption{Ratios of right- and left-handed rotation angles for the set of input parameters shown
in Fig.~\ref{fig:FGspace}.}
\label{fig:rotnangles}
\end{figure}

Since the input parameter
space is multi-dimensional, we have found it convenient to display the regions of interest in a series
of two- and three-dimensional projections.  Figure~\ref{fig:FGspace} shows several such plots
for Level I solutions.  It is clear from the plots that solutions exist for all values
of $\alpha_{\kappa^\prime}$ and $\beta_{23}$.  This observation is consistent with our earlier 
discussion regarding the number of 
input parameters and constraints.  Since we have 11 degrees of freedom and nine essential
constraints~\footnote{Note that the three quark mass ratios are ``redundant'' constraints
in this context, although they are important in our numerical work.},
we expect to have two unconstrained degrees of freedom.
The remainder of the input parameter space is
broken up into several disjoint regions.  These regions actually shrink to a series
of points as $\chi_{(I)}^2$ is reduced to zero.

Our numerical solutions provide an interesting point of contact with
earlier work performed on the SB-LR~\cite{ball1,frere},
where it was pointed out that for each value of $\alpha_{\kappa^\prime}$ there are 64
physically distinct solutions.
In the SB-LR one can associate relative signs with the quarks masses, giving 32
different combinations.  Another factor of ``2'' in that context comes from two possibilities for $\delta_L$ (close
to zero or close to $\pi$).  If our approach is correct and exhaustive, it should be able to find 
all 64 of these solutions.  A good discriminator of the 64 solutions is 
the physically observable
phase $\sigma_d$ (defined in Ref.~\onlinecite{ball1} and also in 
Eq.~(\ref{eq:sigmads}) below), which arises in $B$-$\overline{B}$ oscillations.
Fixing $\alpha_{\kappa^\prime}$ and setting $\beta_{23}=0$ and $\pi$ (recall that $F$ and $G$ are real in the SB-LR)
we have indeed found 64 distinct values for $\sigma_d$, in agreement with
Ref.~\onlinecite{ball1}.
More generally, for any pair of values of the input phases
$\alpha_{\kappa^\prime}$ and $\beta_{23}$ there appear to be 32 distinct solutions.

Figure~\ref{fig:masshist} shows frequency plots of the various quark masses and
rotation angles calculated using the input parameter sets displayed in Fig.~\ref{fig:FGspace}.
The dashed vertical lines in each plot represent the values $y_i^\textrm{\scriptsize exp}\pm \sigma_i$.
The histograms are entirely contained within the dashed lines since our numerical procedure 
ensures that all constraints are satisfied to within 
$\pm 1 \sigma$.  (We have opted for a somewhat
restrictive approach in the present work, but one could easily relax the procedure.)
Note that a narrower distribution indicates that the numerical procedure
had a somewhat ``easier'' time satisfying the given constraint.

Figure~\ref{fig:LRphases} shows the various phases that
characterize the left- and right-handed CKM matrices (see Eqs.~(\ref{eq:vlckm})-(\ref{eq:ktilde}))
for the input parameter sets plotted
in Fig.~\ref{fig:FGspace}.  These plots may be regarded as ``predictions'' of our model
in the sense that the points shown have passed all Level I constraints.
The plots show that all values are possible
for $\delta_L$ (the sole phase in the left-handed CKM matrix) but that
the right-handed phases are typically quite limited by the Level I constraints.
Inclusion of Level II constraints will, of course, further limit the possible values
that the left- and right-handed phases can assume (see Figs.~\ref{fig:LRphasesLevII510} and
\ref{fig:phasesvar} below, for example).
One very interesting result in Fig.~\ref{fig:LRphases} is that $\delta_L$ is very closely
tied to the fundamental phase $\beta_{23}$ in $G$:
\begin{eqnarray}
	\delta_L \approx \beta_{23} + n\pi \pm 0.25 ~\textrm{rad}\; ,
	\label{eq:deltal}
\end{eqnarray}
where $n$ is an integer and where the ``$\pm 0.25$~rad'' indicates the approximate spread of the values
around $\beta_{23} + n\pi$.  This result is consistent with the result
found in Ref.~\onlinecite{ball1}, where $F$ and $G$ were taken to be real matrices
($\beta_{23}=0$ or $\pi$) and where two classes of solutions were found,
one with $|\delta_L|\leq 0.25$ 
and the other with $|\delta_L-\pi|\leq 0.25$.
Also interesting is the approximate equality between $\delta_L$ and $\delta_R$ evident in 
Fig.~\ref{fig:LRphases},
\begin{eqnarray}
	\delta_R\approx \delta_L\pm 0.50~\textrm{rad} \; .
		\label{eq:deltaRapprox}
\end{eqnarray}
In order to understand this result, recall
that in the quasimanifest case (real Higgs VEVs) one has the strict equality
$\delta_R=\delta_L$, while in
the SB-LR (real $F$ and $G$) one has $\delta_R = -\delta_L$.  In the latter case, one has the
additional phenomenological result
that $|\delta_L|\leq 0.25$ or $|\delta_L-\pi|\leq 0.25$, so that $|\delta_R-\delta_L|\leq 0.50$.
Equation~(\ref{eq:deltaRapprox}) may thus be viewed as a marriage of the results
from these two cases: $\delta_R$ is approximately equal to $\delta_L$, as in the
quasimanifest case, but with
a spread of $\pm 0.50$, characteristic of the SB-LR.  (Further discussion along these lines may be found
in Appendix~\ref{sec:appendixb}, where Eqs.~(\ref{eq:deltal}) and (\ref{eq:deltaRapprox}) are derived using an approximate
analytical technique.)
The remaining right-handed phases in Fig.~\ref{fig:LRphases} bear a very close resemblance to
those one obtains in the SB-LR (obtained by
restricting $\beta_{23}$ to the values 0 and $\pi$).
The quasimanifest limit itself is also evident in Fig.~\ref{fig:LRphases}:
the phases $\rho_i$ and $\eta_i$ reduce to $0$ or $\pi$ whenever $\alpha_{\kappa^\prime}=n\pi$.
This behaviour is consistent with the known relation between
the left- and right-handed CKM matrices in the quasimanifest case, namely
$V_{Rij}^{CKM} = \pm V_{Lij}^{CKM}$\cite{deshpande}.

The ratio of right- to left-handed rotation angles (actually, their ``sines'') is shown
in Fig.~\ref{fig:rotnangles}.  
The ratios are identically unity whenever $\alpha_{\kappa^\prime}$ or 
$\beta_{23}$ is equal to $n\pi$, since in these limits our model reduces
to the quasimanifest or pseudomanifest case, respectively.
In a general nonmanifest model the ratios 
are permitted to depart from unity, although our numerical results indicate
that they do not do so by very much.
The largest departure occurs for the ratio
$\sin\theta_{13}^R/\sin\theta_{13}$, which is still typically within 
20$\%$ of unity.  The other two ratios are even closer to unity, with 
$\sin\theta_{23}^R/\sin\theta_{23}$ differing from unity by at most about $0.15\%$.
In Appendix~\ref{sec:appendixb} we explain this intriguing agreement between the left- 
and right-handed rotation angles.  In the case of the ratio of 2-3 angles, for example, the departure
from unity is of order $\lambda^5$ (where $\lambda=0.22$), which is in good agreement with our numerical results.

\section{Level II Constraints: $K$-$\overline{K}$ and $B$-$\overline{B}$ Mixing}
\label{sec:LevII}

The preceding section described the effects of imposing the Level I constraints
on our model.  That analysis was useful in that it served to highlight some of the 
basic properties of the model.  In the present section we describe the
Level II constraints, which are those coming from the neutral $K$ and $B$ systems.
Section~\ref{sec:LevIINum} describes the effects of imposing these additional constraints, an
endeavour that is complicated somewhat by the presence of two new degrees of freedom,
namely the Higgs and $W_2$ masses.  
Table~\ref{tab:chi2} contains a list of the Level II constraints.  These have
been discussed in detail in Ref.~\onlinecite{ball1} and also in Refs.~\onlinecite{ecker85} and
\onlinecite{frere}.
Here we summarize some of the main results in those references.  Note that
we do not attempt to use $\epsilon^\prime$ to place constraints on our results (except in 
Sec.~\ref{sec:SB-LR}, below, where we compare our results with those found by previous authors within the 
SB-LR).

\subsection{Experimental constraints from $\Delta m_{B_d}$ and $\Delta m_{B_s}$}

$\Delta m_{B_d}$ and $\Delta m_{B_s}$ can be quite
sensitive to non-standard contributions in the Left-Right Model~\cite{ecker86,barenboim98,ball1}.
The off-diagonal terms in the mass matrices may be written in terms of a Standard Model
piece and a left-right piece,
\begin{eqnarray}
	M_{12} & = & M_{12}^{SM}+ M_{12}^{LR} \nonumber \\
	       & = & M_{12}^{SM}\left(1 + k_{d(s)} e^{i\sigma_{d(s)}}\right) \; ,	
\end{eqnarray}
where
\begin{eqnarray}
	k_{d(s)} & = & \left|\frac{M_{12}^{LR}}{M_{12}^{SM}}\right|
		\simeq  \left|\frac{V_{R_{tb}}^{CKM} V_{R_{td(s)}}^{CKM*}}
			{V_{L_{tb}}^{CKM} V_{L_{td(s)}}^{CKM*}}\right|
			\left(\frac{B_B^\textrm{\scriptsize{scalar}}}{B_B}\right)
			\left\{ \left(\frac{7~\textrm{TeV}}{M_H}\right)^2  \right. \nonumber \\
	& &	\left. ~~~~~~~~+ \eta_2^{LR} \left(\frac{1.6~\textrm{TeV}}{M_2}\right)^2
		\left[ 0.051 -0.013 \ln \left( \frac{1.6~\textrm{TeV}}{M_2}\right)^2 \right]
				\right\} \; 
		\label{eq:kds}
\end{eqnarray}
and
\begin{eqnarray}
	\sigma_{d(s)} = \arg\left(-\frac{V_{R_{tb}}^{CKM} V_{R_{td(s)}}^{CKM*}}
			{V_{L_{tb}}^{CKM} V_{L_{td(s)}}^{CKM*}}\right) \; ,
	\label{eq:sigmads}
\end{eqnarray}
with $B_B^\textrm{\scriptsize{scalar}}/B_B\simeq 1.2$ and $\eta_2^{LR}\simeq 1.7$~\cite{ball1}.
The expression for $k_{d(s)}$ is an approximation that is accurate to about $5\%$ for
$M_2>1.4$~TeV and $M_H>7$~TeV.
The full expression for $M_{12}^{SM}$ may be found in Ref.~\onlinecite{ball1}.  Our
expression for $\sigma_{d(s)}$ is identical to that found in Ref.~\onlinecite{ball1}, while
that for $k_{d(s)}$ differs in that it contains a ratio of right- and left-handed CKM matrix elements.
This ratio is equal to unity in the spontaneously-broken model considered in Ref.~\onlinecite{ball1}.

The above results may be used to solve for the $B$-$\overline{B}$ mass differences, since
\begin{eqnarray}
	\Delta m_{B_{d(s)}} = 2 \left| M_{12}\right| \; .
\end{eqnarray}
In the case of $B_d$, the mass difference is quite well known experimentally.  Nevertheless,
various theoretical uncertainties relax the bound somewhat, leading
to the following constraint~\cite{ball1},
\begin{eqnarray}
	\left|\left(V_{L_{tb}}^{CKM} V_{L_{td}}^{CKM*}\right)^2
		\left(1 + k_{d} e^{i\sigma_{d}}\right)\right| = \left(6.7 \pm 2.7\right)\times 10^{-5} \; .
\end{eqnarray}
In terms of $\Delta m_{B_d}$ itself, the above range corresponds to 
$\Delta m_{B_d} = 0.472 \pm 0.190~\textrm{ps}^{-1}$, as is quoted in Table~\ref{tab:inputranges}.
In the case of $B_s$ there is only an experimental lower bound on the mass difference,
$\Delta m_{B_s}\geq 15.0$~ps$^{-1}$~\cite{ciuchini,atwoodsoni,hocker,LEPwg}.
When comparing to theoretical expectations, the lower bound is usually expressed as a ratio, since this 
tends to decrease the theoretical uncertainty.  We enforce the following bound, 
\begin{eqnarray}
	\frac{\Delta m_{B_s}}{\Delta m_{B_d}} = 1.31\times
		\left|\frac{\left(V_{L_{tb}}^{CKM} V_{L_{ts}}^{CKM*}\right)^2
		\left(1 + k_{s} e^{i\sigma_{s}}\right)}
			{\left(V_{L_{tb}}^{CKM} V_{L_{td}}^{CKM*}\right)^2
		\left(1 + k_{d} e^{i\sigma_{d}}\right)}\right|
		\geq 27.2 \; .
	\label{eq:mbsbound}
\end{eqnarray}
This bound takes into account theoretical uncertainties and is slightly modified from that given in 
Ref.~\onlinecite{ball1}.
The $\Delta m_{B_s}$ constraint is enforced in a different manner than most other constraints, since
it is only included as a ``cut'' after a potential
solution has been identified (i.e., $\Delta m_{B_s}$ is not included in the evaluation of $\chi^2_{(II)}$).

\subsection{Experimental constraints from $\epsilon_K$ and $\Delta m_K$}

The $K$-$\overline{K}$ system has long played an important role in constraining the 
Left-Right Model.  $\Delta m_K$ puts a lower bound of about 1.6~TeV on the mass
of $W_2$~\cite{bbs}, while $\epsilon_K$ can in principle put a lower bound of about 
50~TeV on the Higgs mass~\cite{pospelov}.  
This latter bound is due to the presence of a tree-level FCNH contribution
to $\epsilon_K$.  As we shall see in Sec.~\ref{sec:LevIINum}, a detailed numerical treatment of
our model indicates that the experimental bounds may be satisfied with Higgs masses as low as about 7~TeV.
Both $\epsilon_K$ and $\Delta m_K$
are defined in terms of $M_{12}$, the off-diagonal term in the $K$-$\overline{K}$
mass matrix~\cite{ball1},
\begin{eqnarray}
	\epsilon_K & = & \frac{e^{i\pi/4}}{\sqrt{2}}\left(\frac{\textrm{Im}(M_{12})}{\Delta m_K^\textrm{\scriptsize exp}}
		+\xi_0\right)
		\label{eq:epsk} \\
	\Delta m_K & = & 2 \; \textrm{Re}\left(M_{12}\right) 
		\label{eq:deltamk}\; ,
\end{eqnarray}
where
\begin{eqnarray}
	\xi_0 = \frac{\textrm{Im}a_0}{\textrm{Re}a_0}, 
		~~~~~~~~a_0^* = \langle \pi\pi (I=0)|-i{\cal H}_\textrm{\scriptsize eff}^{|\Delta S |= 1}|
		\overline{K}^0\rangle_\textrm{\scriptsize weak}.
\end{eqnarray}
$\Delta m_K$ suffers from relatively large theoretical uncertainties due to 
long-distance contributions, so we follow the usual practice of using the
experimental value for $\Delta m_K$ in Eq.~(\ref{eq:epsk}) rather than the theoretical (short
distance) expression obtainable from $M_{12}$.
The term proportional to $\xi_0$ in Eq.~(\ref{eq:epsk}) is also subject to considerable theoretical
uncertainties.  Within the SM these uncertainties do not pose any particular difficulties because the 
contribution due to this term is quite small and may safely be neglected.  Such is not necessarily the case
within the Left-Right Model, where the contribution due to $\xi_0$ can be of order $30\%$ for
$M_2=1.6$~TeV~\cite{ecker85,frere, ball1}.  We follow Ref.~\onlinecite{ball1} in ignoring
the $\xi_0$ contribution to $\epsilon_K$ and in taking its effect into account through a theoretical
uncertainty.

Reference~\onlinecite{ecker85} contains a thorough discussion of the various contributions
to $M_{12}$ within the Left-Right Model.  
Interestingly, the sum of box diagrams is not itself gauge invariant in the Left-Right 
Model~\cite{housoni,chang}.  Nevertheless,
the diagrams restoring gauge invariance give
very small contributions in the 't Hooft-Feynman gauge, and can safely be ignored
while working in that gauge.  Similarly, several of the box diagrams generically
give quite small contributions and can be ignored, leaving a total of five terms in the theoretical
expression for $M_{12}$~\cite{ecker85},
\begin{eqnarray}
	M_{12} = M_{12}^{SM}+M_{12}^{FCNH}+M_{12}^{W_1W_2}+M_{12}^{S_1W_2} +M_{12}^{W_1\Phi^\pm} \; ,
		\label{eq:M12delmk}
\end{eqnarray}
where the first term is the usual SM contribution, the second corresponds to the
tree-level FCNH contribution, and $S_1$ and $\Phi^\pm$ 
refer to one of the unphysical scalars and to the physical charged Higgs, respectively.
(Recall that, for simplicity, all non-standard Higgs bosons are taken to have the same mass in the present
work.)  Explicit expressions for the various terms may be found in Refs.~\onlinecite{ecker85}
and~\onlinecite{bbl} and are not included here~\footnote{We use 
the NLO results in Ref.~\onlinecite{bbl}
for the SM piece and the LO results in 
Ref.~\onlinecite{ecker85} for the left-right pieces.}.

Table~\ref{tab:5pieces} shows a numerical evaluation of 
the five short-distance contributions to $\epsilon_K$ and
$\Delta m_K$ for a particular Level I solution and gives a rough indication 
of how the various terms scale with increasing Higgs and $W_2$ masses.  (There
is nothing particularly ``special'' about this data point other than that it
happens to give a SM contribution that is
close to the known experimental value.)  One of the most striking features of
the table is the very large tree-level FCNH contribution to $\epsilon_K$.
Pospelov studied this contribution a few
years ago in the case of manifest left-right symmetry and
concluded that the corresponding Higgs boson would be required to 
have a mass in excess of 50 TeV~\cite{pospelov}.
The appropriateness of this bound is evident in the last row of the table, where the troublesome
term is seen to reach a manageable size once $M_H\agt 50$~TeV.  
Having said this, let us note that we are in fact able to find
complete Level II solutions with Higgs masses of order 7~TeV.  Such solutions
do require a certain amount of ``fine-tuning,'' but they exist nonetheless.

\begin{table*}[t]
\caption{Short-distance contributions to $\epsilon_K$ and $\Delta m_K$ for a particular
data point that satisfies all Level I (but not Level II) constraints.  The left- and right-handed phases
for this Level I solution are $\delta_L = 0.9973$ and $(\rho_1, \rho_2, \rho_3, \eta_2, \eta_3,
\delta_R) = (2.979, \;0.1758, \;3.168, \;3.134, \;4.783, \;0.5801)$ and the sines of
the right-handed rotation angles are $(\sin\theta_{12}^R,\sin\theta_{23}^R,\sin\theta_{13}^R) = 
(0.2254,\; 0.0396, \; 0.00279)$.  We have set
$\Lambda_3 = 0.350$~GeV, $\mu = 1.0$~GeV and $B_K = 0.86$.
\vspace{.2in}}
\label{tab:5pieces}
\begin{tabular}{|c|c|ccccc|ccccc|}\hline\hline
	$M_2$ & $M_H$ &
	\multicolumn{5}{c|}{$\epsilon_K^{SD}/\epsilon_K^\textrm{\scriptsize exp}$} &
	\multicolumn{5}{c|}{$\Delta m_K^{SD}/\Delta m_K^\textrm{\scriptsize exp}$} \\ \hline
	(TeV) & (TeV) & SM & FCNH & $W_1 W_2$ & $S_1 W_2$ & $W_1 \Phi^\pm$ &
		SM & FCNH & $W_1 W_2$ & $S_1 W_2$ & $W_1 \Phi^\pm$ \\ \hline
	1.6 & 5 & 0.798 & -31.9 & -0.224 & -1.61 & -0.250 & 0.747 & 0.770 & 0.608 & 0.071 & 0.009 \\
	5 & 5   & 0.798 & -31.9 & -0.023 & -0.249 & -0.250 & 0.747 & 0.770 & 0.064 & 0.009 & 0.009 \\
	5 & 10  & 0.798 & -7.98 & -0.023 & -0.249 & -0.075 & 0.747 & 0.193 & 0.064 & 0.009 & 0.003 \\
	5 & 50  & 0.798 & -0.319 & -0.023 & -0.249 & -0.004 & 0.747 & 0.008 & 0.064 & 0.009 & 0.0001 \\
			\hline\hline
\end{tabular}
\end{table*}

The left-right contributions to $\epsilon_K$ in Ref.~\onlinecite{ecker85} are 
only accurate to LO and display a relatively strong
dependence on the low-energy QCD scales 
$\mu$ and $\Lambda_3$.  The SM piece, while evaluated
to NLO in the present work, is also subject to uncertainty due to the kaon bag parameter, $B_K$.
In order to investigate the effects of these uncertainties, we have examined $\epsilon_K$ predictions
for a set of data points that passed the Level I constraints.  We combined the data points
with various Higgs and $W_2$ mass combinations and evaluated $\epsilon_K$ 
taking $\mu$, $\Lambda_3$ and $B_K$ in the ranges
$\mu = 1.0\pm 0.2$~GeV, $\Lambda_3=0.350\pm 0.100$~GeV and $B_K=0.86\pm 0.15$.  The resulting
spread of $\epsilon_K$ values typically fell within $20-30\%$ of the mean.  Rather than allowing these
three parameters to vary in our numerical work, we have fixed them to the ``central'' values
($\mu = 1.0$~GeV, $\Lambda_3=0.350$~GeV and $B_K=0.86$) and have assigned a $20\%$ theoretical uncertainty
to $\epsilon_K$.  Using the value
$\epsilon_K = 2.28\times 10^{-3}$~\cite{pdg2000}
and ignoring the small experimental uncertainty,
we obtain~\footnote{We have used $N_f=3$ when evaluating
the expressions in Ref.~\onlinecite{ecker85}.
In some of the expressions (such as in the treel-level FCNH contribution)
one could in principle evolve the Wilson coefficients in several steps rather than
all at once,
but it is not clear that this would be appropriate in some of the other expressions.}
\begin{eqnarray}
	\epsilon_K = \left[2.28 \pm \left(0.46+3.5 \times
		(1.0~\textrm{TeV}/M_2)^2\right)\right]\times 10^{-3} \; ,
\end{eqnarray}
where the first term in the uncertainty is due to 
uncertainties in $\mu$, $\Lambda_3$ and $B_K$.  The second term in the uncertainty is
due to our neglect of the $\xi_0$ term in Eq.~(\ref{eq:epsk}) and is
taken from Ref.~\onlinecite{ball1}.

Theoretical expressions for $\Delta m_K$ involve large uncertainties due
to long distance contributions, even within the context of the SM.  The SM
calculation of the short-distance contribution to $\Delta m_K$ gives roughly
$70\%$ of the known experimental value (see Table~\ref{tab:5pieces}).  Within the SM,
the remaining $30\%$ is thought to be due to long-distance effects.  It is not clear how one
might best use $\Delta m_K$ to place constraints on non-standard physics, since the
long-distance contributions are somewhat unknown.  We follow previous authors and constrain
new contributions to be at most as large as $\Delta m_K^\textrm{\scriptsize{exp}}$ 
itself.  Our constraints are
\begin{eqnarray}
	-1\leq 2\; \textrm{Re}(M_{12}^{LR})/\Delta m_K^\textrm{\scriptsize{exp}}\le 1
		~~~~\textrm{and}~~~~ \textrm{Re}(M_{12})>0 \; .
\end{eqnarray}
The $\Delta m_K$ constraints are implemented as cuts and are not used in the evaluation
of $\chi^2_{(II)}$.

\subsection{Experimental constraint from $B\to \psi K_S$}

Recent measurements of $\sin 2\beta_{CKM}^\textrm{\scriptsize eff}$ 
by the BABAR and BELLE collaborations yield the values
$0.59\pm 0.14\pm 0.05$ and $0.99\pm 0.14 \pm 0.06$, respectively~\cite{babar2001,belle2001}.
Taking the weighted average yields the value
$\sin 2\beta_{CKM}^\textrm{\scriptsize eff}=0.79\pm 0.11$,
which is consistent with the slightly older CDF measurment~\cite{cdf2000}.  This experimental
value actually acts to constrain both the CKM angle ``$\beta_{CKM}$,'' as well as non-standard
effects coming from $K$-$\overline{K}$ and $B$-$\overline{B}$ mixing.  The full theoretical
expression is given by~\cite{ball1}
\begin{eqnarray}
	\sin 2\beta_{CKM}^\textrm{\scriptsize eff} = \sin\left[2 \beta_{CKM} +
		\arg\left(1+k_de^{i\sigma_d}\right)-\arg\left(1+\frac{M_{12}^{K,LR}}
			{M_{12}^{K,SM}}\right)\right] \; ,
\end{eqnarray}
where
\begin{eqnarray}
	\beta_{CKM} = \arg\left(-\frac{V_{L_{cd}}^{CKM} V_{L_{cb}}^{CKM*}}
			{V_{L_{td}}^{CKM} V_{L_{tb}}^{CKM*}} \right) \; .
\end{eqnarray}
When employing $\sin 2\beta_{CKM}^\textrm{\scriptsize eff}$ as a constraint we 
take into account the experimental uncertainty, but do not include any additional theoretical
uncertainty.

\section{Level II Solutions: Numerical Results}

\label{sec:LevIINum}

In this section we employ all the experimental constraints in Table~\ref{tab:chi2} in order
to search the parameter space of the model for Level II solutions.  Section~\ref{sec:SB-LR} contains
a study of our model in the pseudomanifest limit ($F$ and $G$ real), in which case our model
reduces to the SB-LR~\cite{ball1}.  In Secs.~\ref{sec:5-10} and \ref{sec:variablemasses} we perform
two case studies.  In the first we fix the $W_2$ and Higgs masses 
to be 5 and 10 TeV, respectively, while in the second we
allow the masses to vary over prescribed ranges.

\subsection{Comparison with results in the SB-LR}
\label{sec:SB-LR}

\begin{figure}
\resizebox{\textwidth}{!}{\includegraphics{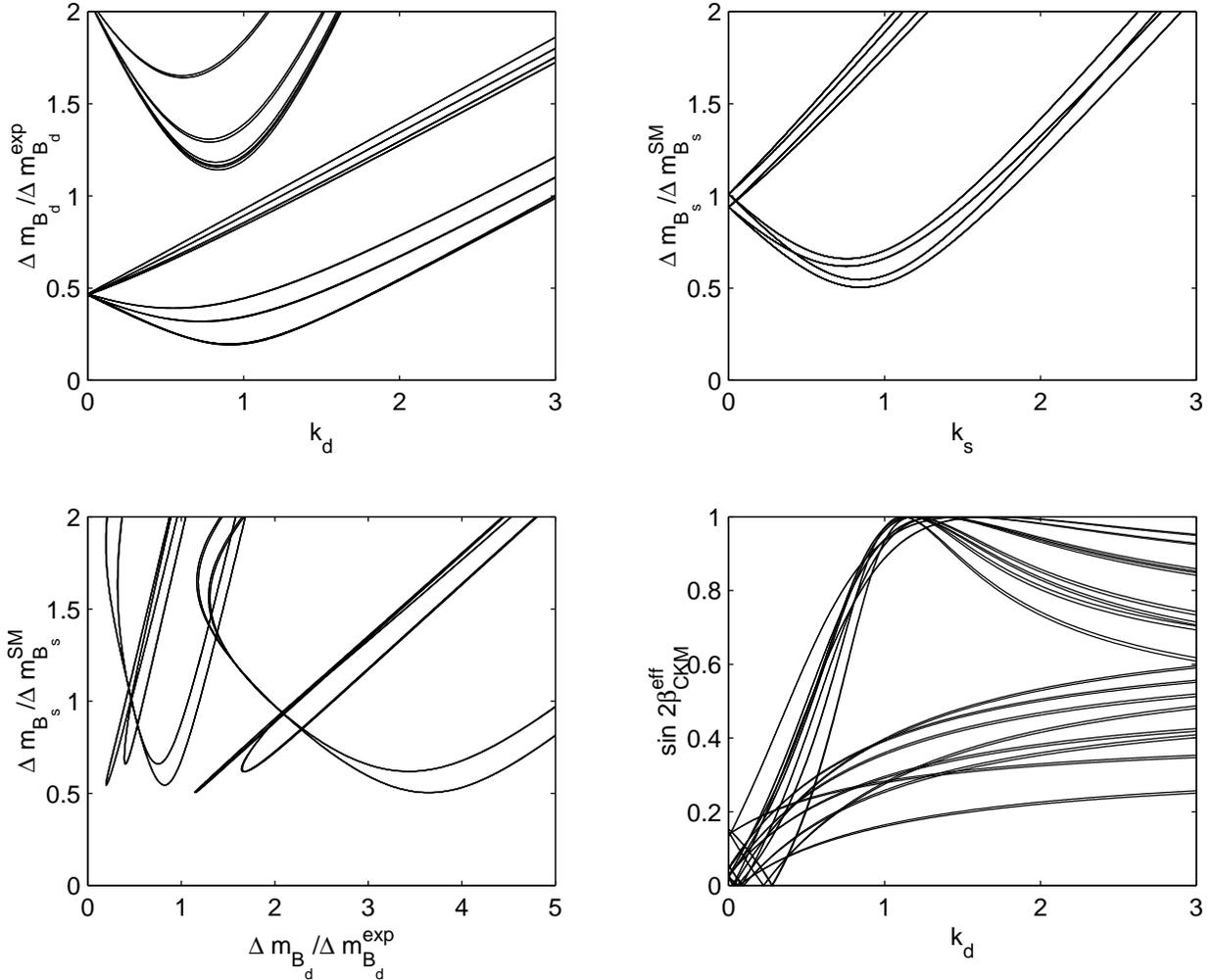}}
\caption{Reproduction of some results obtained for the SB-LR in Ref.~\onlinecite{ball1}.
The above plots were obtained by setting $\beta_{23}=0,\pi$ and
$\alpha_{\kappa^\prime} = \sin^{-1} \left((1-r^2)\tan\beta/(2r)\right)$, with $r=|\kappa^\prime/\kappa|$
and $\beta=0.02$,
in the notation of Ball et al.  The plots
may be compared with the $\beta=0.02$ case in each of Figs.~4-7 in
Ref.~\onlinecite{ball1}.  Note that $k_s=k_d$ in the SB-LR, since the ratio
of CKM matrix elements in Eq.~(\ref{eq:kds}) is equal to unity in
that case.  Note also that the above plot of 
$\sin 2\beta_{CKM}^\textrm{\scriptsize eff}$ neglects the $K$-$\overline{K}$
mixing contribution, as does the analogous plot in Ref.~\onlinecite{ball1}.
Only non-negative values of $\sin 2\beta_{CKM}^\textrm{\scriptsize eff}$ are shown.}
\label{fig:delmb}
\end{figure}

We begin by using our method to rederive some
of the results obtained in the SB-LR~\cite{ball1},
since this serves as a useful check of our method.  Figure~\ref{fig:delmb}
shows plots of $\Delta m_{B_d}/\Delta m_{B_d}^\textrm{\scriptsize exp}$,
$\Delta m_{B_s}/\Delta m_{B_s}^{SM}$ and 
$\sin 2\beta_{CKM}^\textrm{\scriptsize eff}$ for a set of Level I solutions
generated for a particular value of the
phase $\alpha_{\kappa^\prime}$, where~\cite{ball1}
\begin{eqnarray}
	\frac{\Delta m_{B_d}}{\Delta m_{B_d}^\textrm{\scriptsize exp}} & = &	
		\frac{|(V_{L_{tb}}^{CKM} V_{L_{td}}^{CKM*})^2
	\left(1 + k_{d} e^{i\sigma_{d}}\right)|}{\left(6.7 \times 10^{-5}\right)} \\
	\frac{\Delta m_{B_s}}{\Delta m_{B_s}^{SM}} & = &
		\frac{|(V_{L_{tb}}^{CKM} V_{L_{ts}}^{CKM*})^2
	\left(1 + k_{s} e^{i\sigma_{s}}\right)|}{0.039^2}  \; .
\end{eqnarray}
In each case the plot was obtained by allowing $\beta_{23}$ to take
on the values $0$ and $\pi$, since our model reduces to the SB-LR for
these values of $\beta_{23}$.  In order to reproduce the results
of Ball et al. more precisely, we have enforced the rather
stringent bound $\chi_{(I)}^2<2\times 10^{-6}$ for this particular
set of points.  Comparison of this figure with Figs.~4-7 in Ref.~\onlinecite{ball1}
shows quite good agreement of our results with those obtained there.
There are, however, two differences between our results and those in Ref.~\onlinecite{ball1}.
In the first place, the overall shapes of the plots are slightly different.  This difference is due
to a slightly different choice of quark masses and left-handed rotation angles.  A second difference
concerns the number of lines evident in the plots.  In some places where we appear to have a single
line (or several very closely spaced lines), Ball et al. have several lines.
One possibility would be that our method is actually missing some solutions.  We do not believe this to be
the case, however, since our evaluation of $\sigma_d$ for this case shows 64 distinct values. 
The reason for this difference is not clear to us.

\begin{figure}
\resizebox{\textwidth}{!}{\includegraphics{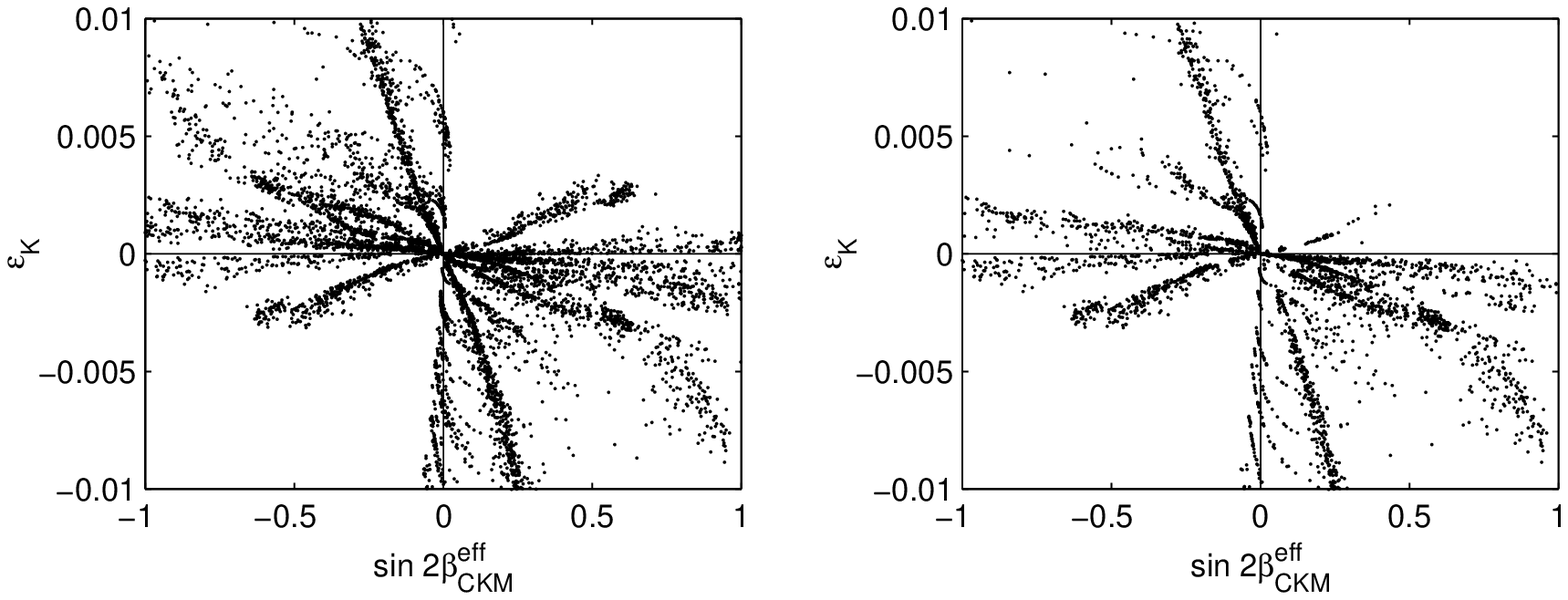}}
\caption{Plot of $\epsilon_K$ versus $\sin 2\beta_{CKM}^\textrm{\scriptsize eff}$
in the SB-LR.  The plot on the left satisfies all $\Delta m$-type bounds
(as described in the text).  The plot on the right satisfies the additional
constraint $e^{-i\pi/4}\epsilon^\prime>0$ and may be compared with Fig.~9 in Ref.~\onlinecite{ball1}.}
\label{fig:epsksin2b}
\end{figure}

We may also draw a comparison with Fig.~9 in Ref.~\onlinecite{ball1}, which shows a plot
of $\epsilon_K$ versus $\sin2\beta_{CKM}^\textrm{\scriptsize eff}$ for a range
of values for $M_2$ and $M_H$.  Figure~\ref{fig:epsksin2b} shows two plots
of $\epsilon_K$ versus $\sin2\beta_{CKM}^\textrm{\scriptsize eff}$ for the
case when $\beta_{23}=0,\pi$.  For each of the plots
a set of Level I solutions is combined with many pairs of
$W_2$ and Higgs masses.  The points shown have passed the $\Delta m_{B_d}$,
$\Delta m_{B_s}$  and $\Delta m_{K}$ bounds in Table~\ref{tab:chi2}~\footnote{When comparing
with the SB-LR we began with Level I solutions and enforced all Level II constraints as 
``cuts.''} and correspond to masses in the range $2<M_2<16$~TeV and $4<M_H<18$~TeV.
The plot on the right has passed an additional cut on 
$\epsilon^\prime=\epsilon^\prime_{SM}+\epsilon^\prime_{LR}$.
In evaluating $\epsilon^\prime$, we have taken the SM piece from
Ref.~\onlinecite{bbl}.  For the LR piece we have used the following
expression~\cite{frere},
\begin{eqnarray}
	\epsilon^\prime_{LR} & \simeq & e^{i\pi/4}\times 10^{-2} \times\left\{
		\left[6.8 \left[\frac{\alpha_s(\mu^2)}{\alpha_s(M_2^2)}\right]^{-2/b}
		-0.30 \left[\frac{\alpha_s(\mu^2)}{\alpha_s(M_2^2)}\right]^{4/b}\right]
		\frac{M_1^2}{M_2^2}\sin(-\eta_2) \right. \nonumber \\
		& &  ~~~~~~~~~~~~~~~~~~~~~~+102\zeta \left[\sin(\alpha_{\kappa^\prime}+\rho_1-\eta_2)+
			\sin(\alpha_{\kappa^\prime}+\rho_1)\right] \nonumber \\	
		& &  ~~~~~~~~~~~~~~~~~~~~~~\left. -9.6\zeta \left[\sin(\alpha_{\kappa^\prime}+\rho_2)+
			\sin(\alpha_{\kappa^\prime}+\rho_2-\eta_2)\right]
			\frac{}{} \!\! \right\} \!,
\end{eqnarray}
where 
\begin{eqnarray}
	\zeta=\frac{2r}{1+r^2}\left(\frac{M_1}{M_2}\right)^2 \; ,
\end{eqnarray}
with $r=|\kappa^\prime/\kappa|$ and $b=11-2N_f/3$.  
The above expression for $\epsilon^\prime_{LR}$
has been modified from that in Ref.~\onlinecite{frere} in order
to account for a difference in gauge choice.

Comparison of Fig.~\ref{fig:epsksin2b} in the present work with Fig.~9 in Ref.~\onlinecite{ball1}
shows reasonable agreement between the two plots, but there are a few
differences.  In particular, while Ball et al. 
find no solutions near the experimental values for $\epsilon_K$ and 
$\sin 2\beta_{CKM}^\textrm{\scriptsize eff}$, we do find solutions that are somewhat close to these
values.  Also, while Ball et al. have very few points in the third quadrant, our plot
shows a fairly prominent band in this region.
It is unclear to us why our results differ from those found by Ball et al., particularly
given the good agreement between our Fig.~\ref{fig:delmb} and their Figs.~4-7.
It is possible that the discrepancy is due to small differences
in our evaluation of the expressions for $\epsilon_K$ or $\Delta m_K$ 
in Ref.~\onlinecite{ecker85} or in our evaluation and application of the $\epsilon^\prime$ constraint.
We should emphasize that, due to the significant theoretical uncertainties involved 
in the calculation of $\epsilon^\prime$, we do not use $\epsilon^\prime$ in the remainder of our analysis.
We have only discussed it in the present subsection in order to facilitate a
comparison with the work in Ref.~\onlinecite{ball1}.

\subsection{A case study: $\mathbf{{\textit M_2}=5}$~TeV and $\mathbf{{\textit M_H}=10}$~TeV}
\label{sec:5-10}

\begin{figure}
\resizebox{\textwidth}{!}{\includegraphics{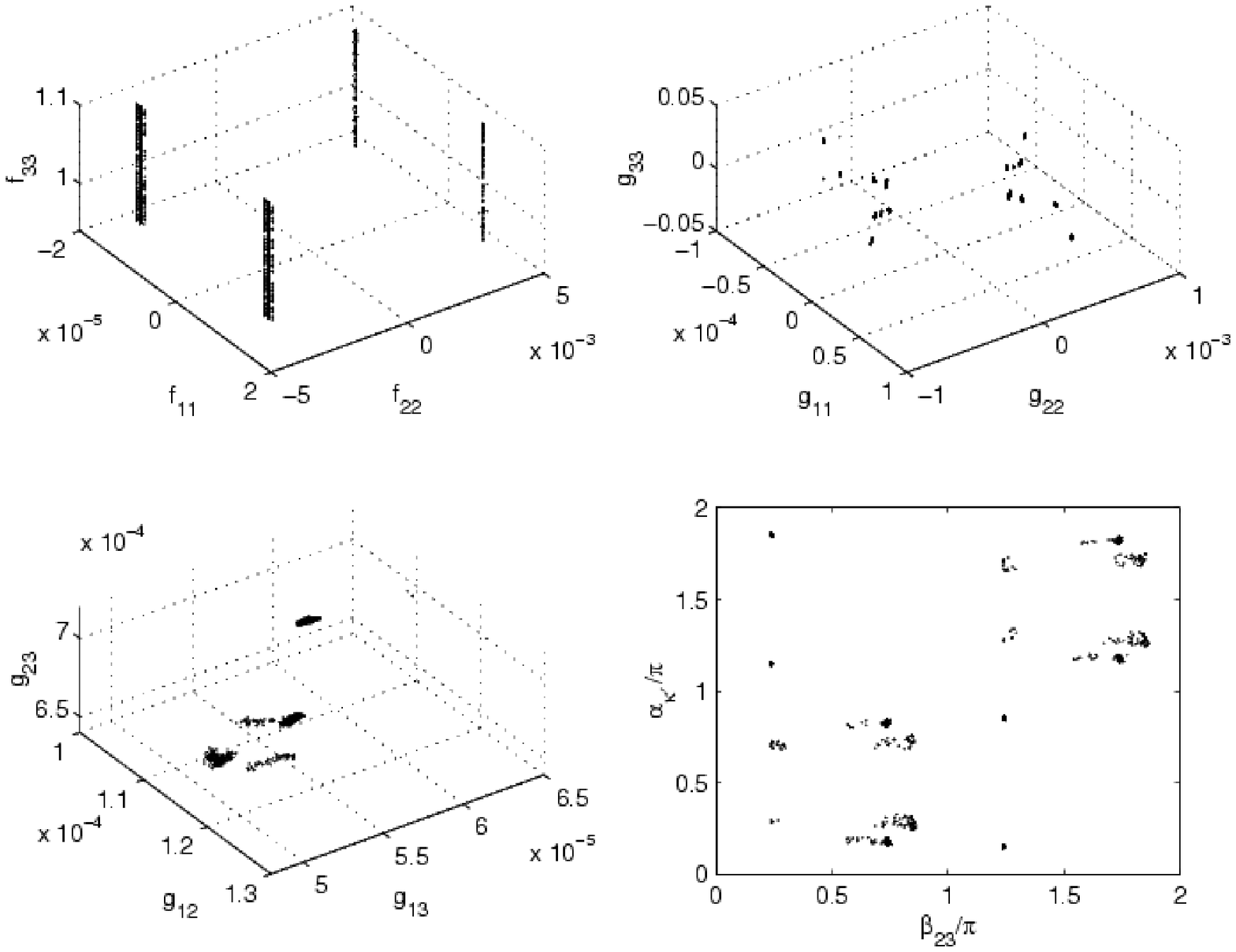}}
\caption{Regions of the input parameter space that 
yield Level II solutions when $M_2=5$~TeV and $M_H=10$~TeV.}
\label{fig:fgspaceLevII510}
\end{figure}
\begin{figure}
\resizebox{\textwidth}{!}{\includegraphics{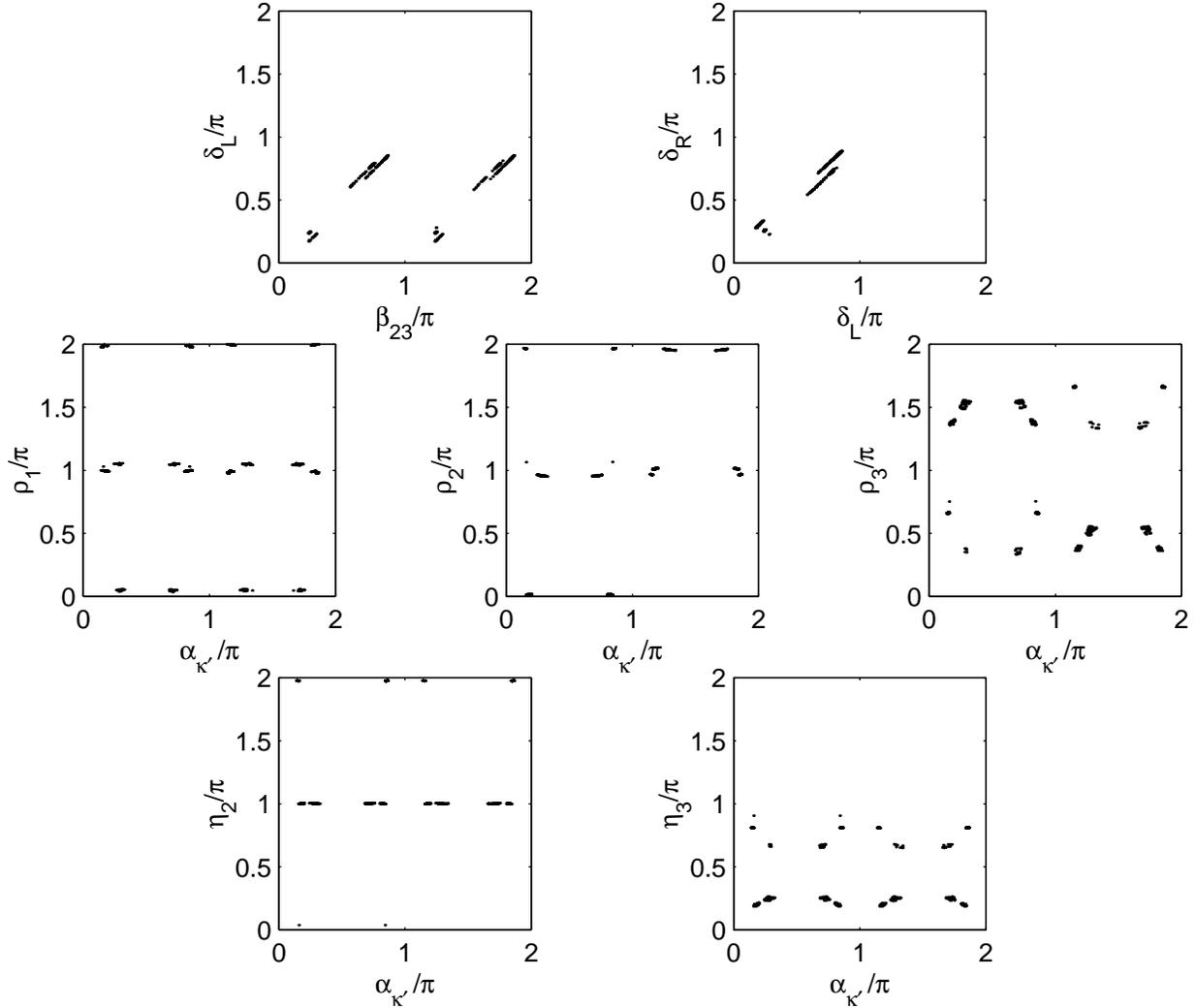}}
\caption{Left- and right-handed CKM phases for the set of input parameters in 
Fig.~\ref{fig:fgspaceLevII510}.  These are Level II solutions with $M_2=5$~TeV and $M_H=10$~TeV.}
\label{fig:LRphasesLevII510}
\end{figure}
\begin{figure}
\resizebox{\textwidth}{!}{\includegraphics{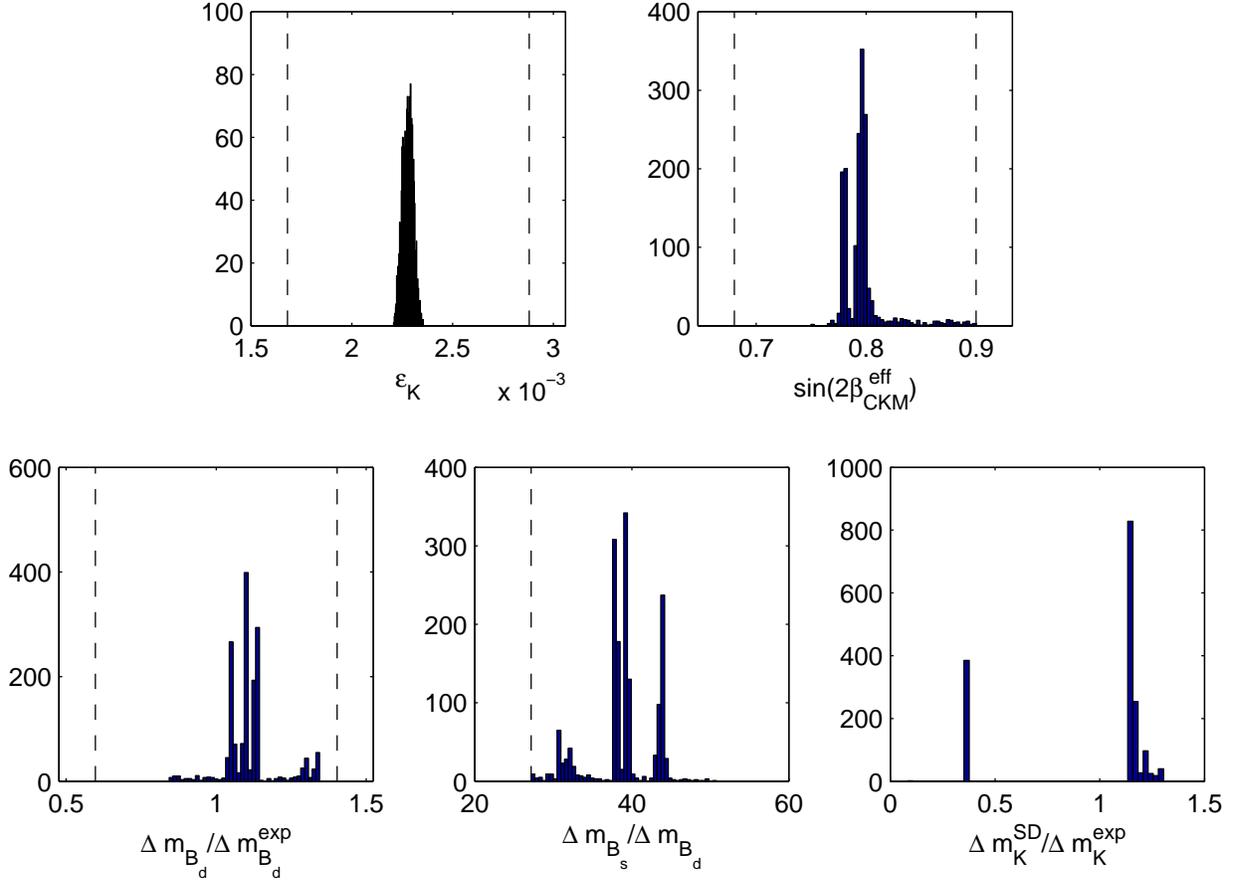}}
\caption{Frequency plots of the five Level II constraints for the data
set in Fig.~\ref{fig:fgspaceLevII510}.  The dashed vertical lines for 
$\epsilon_K$, $\sin 2\beta_{CKM}^\textrm{\scriptsize eff}$
and $\Delta m_{B_d}$ indicate the theoretical and/or experimental uncertainties.
The dashed line for $\Delta m_{B_s}$ indicates the lower bound on
$\Delta m_{B_s}/\Delta m_{B_d}$ (see Eq.~(\ref{eq:mbsbound})). 
The data in the plot are Level II solutions with $M_2=5$~TeV and $M_H=10$~TeV.}
\label{fig:fiveLevII510}
\end{figure}
\begin{figure}
\resizebox{\textwidth}{!}{\includegraphics{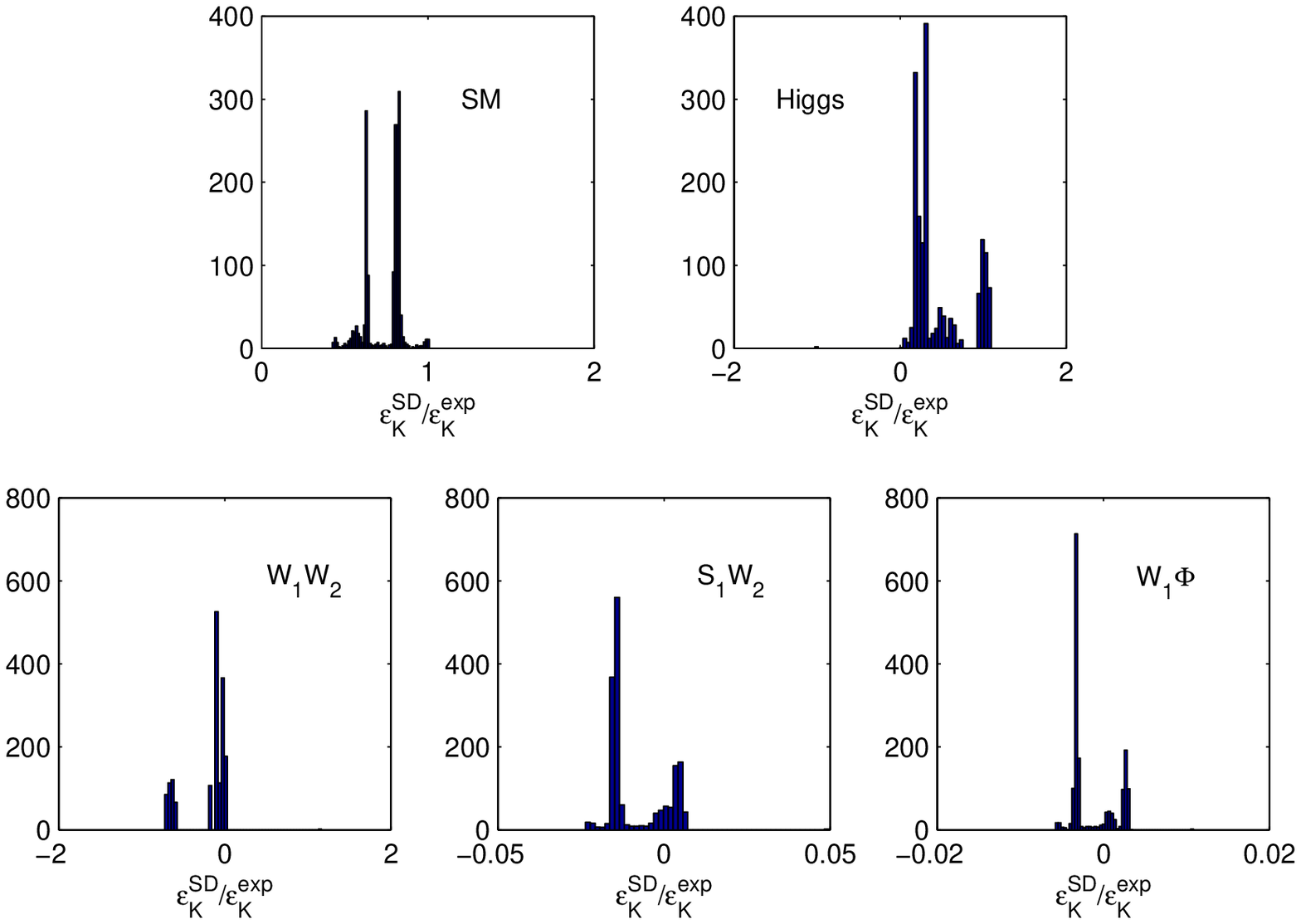}}
\caption{Frequency plots of the five short-distance contributions to $\epsilon_K$ for the data
set in Fig.~\ref{fig:fgspaceLevII510}.  Note the different scales on the horizontal axes.
The data in the plot are Level II solutions with $M_2=5$~TeV and $M_H=10$~TeV.}
\label{fig:epskhist510}
\end{figure}

We turn now to a case study for a particular pair of $W_2$ and Higgs masses, choosing 
$M_2=5$~TeV and $M_H=10$~TeV.  Figures~\ref{fig:fgspaceLevII510}-\ref{fig:epskhist510}
show our numerical results for Level II solutions in this case.  
Figure~\ref{fig:fgspaceLevII510} contains a plot of the input parameter space showing points
that satisfy the Level II constraints when $M_2=5$~TeV and $M_H=10$~TeV.  This figure
may be compared with Fig.~\ref{fig:FGspace}, which shows a set of points that pass 
only the Level I constraints.  Note that there are severe constraints on the possible values for $\alpha_{\kappa^\prime}$ and
$\beta_{23}$.  In particular, both the quasimanifest ($\alpha_{\kappa^\prime}=n\pi$) and pseudomanifest ($\beta_{23}=n\pi$) limits
would appear to be ruled out for this particular pair
of values for the masses $M_2$ and $M_H$.
Figure~\ref{fig:LRphasesLevII510} shows
the left- and right-handed CKM phases for the set of input parameters in 
Fig.~\ref{fig:fgspaceLevII510}.  This figure
may similarly be compared with Fig.~\ref{fig:LRphases}.  The Level II constraints 
rule out most of the possible values that the right-handed phases could in principle assume.
Figure~\ref{fig:fiveLevII510} shows frequency distributions for the five quantities
yielding the Level II constraints.  The dashed vertical lines indicate experimental
and/or theoretical bounds in each case.  The histogram plot in Fig.~\ref{fig:epskhist510}
shows the frequency distributions for each of the five short-distance contributions to 
$\epsilon_K$.  The SM, FCNH and $W_1$-$W_2$ contributions can all be quite significant.
Some further investigation has also shown a relatively strong anti-correlation
between the FCNH and $W_1$-$W_2$ contributions that yield Level II solutions: when 
the contributions are large they tend to be of opposite sign.

\subsection{The generic case: variable masses}
\label{sec:variablemasses}

\begin{figure}
\resizebox{11cm}{!}{\includegraphics{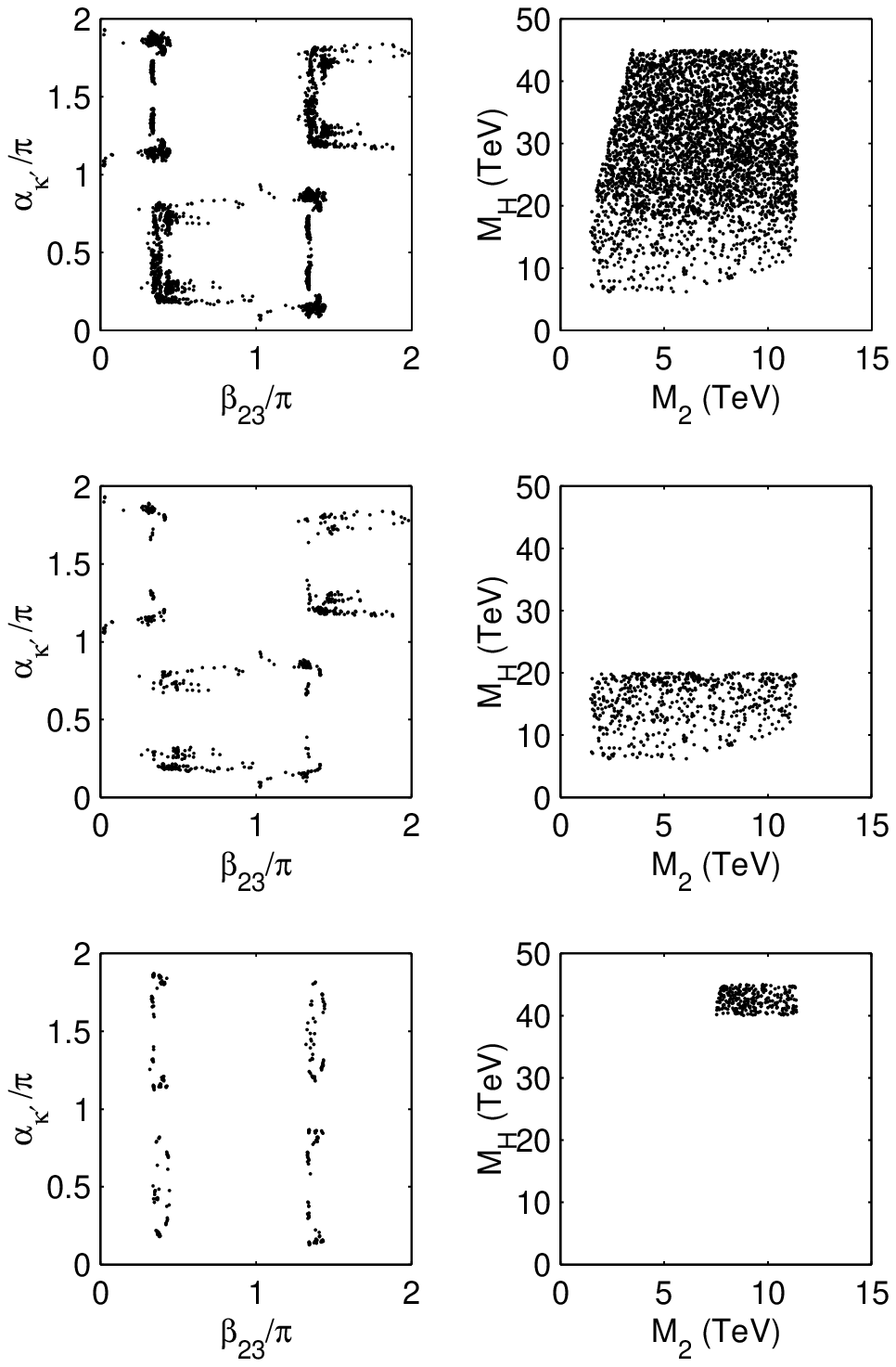}}
\caption{Three pairs of plots showing Level II solutions
for the variable mass case.  The plots on the right indicate the mass
ranges under consideration and the plots on the left show the
corresponding values of $\alpha_{\kappa^\prime}$ and $\beta_{23}$.
The middle and lower pairs of plots 
show subsets of the data contained in the upper pair of plots.}
\label{fig:solnsLevIIvar}
\end{figure}
\begin{figure}
\resizebox{\textwidth}{!}{\includegraphics{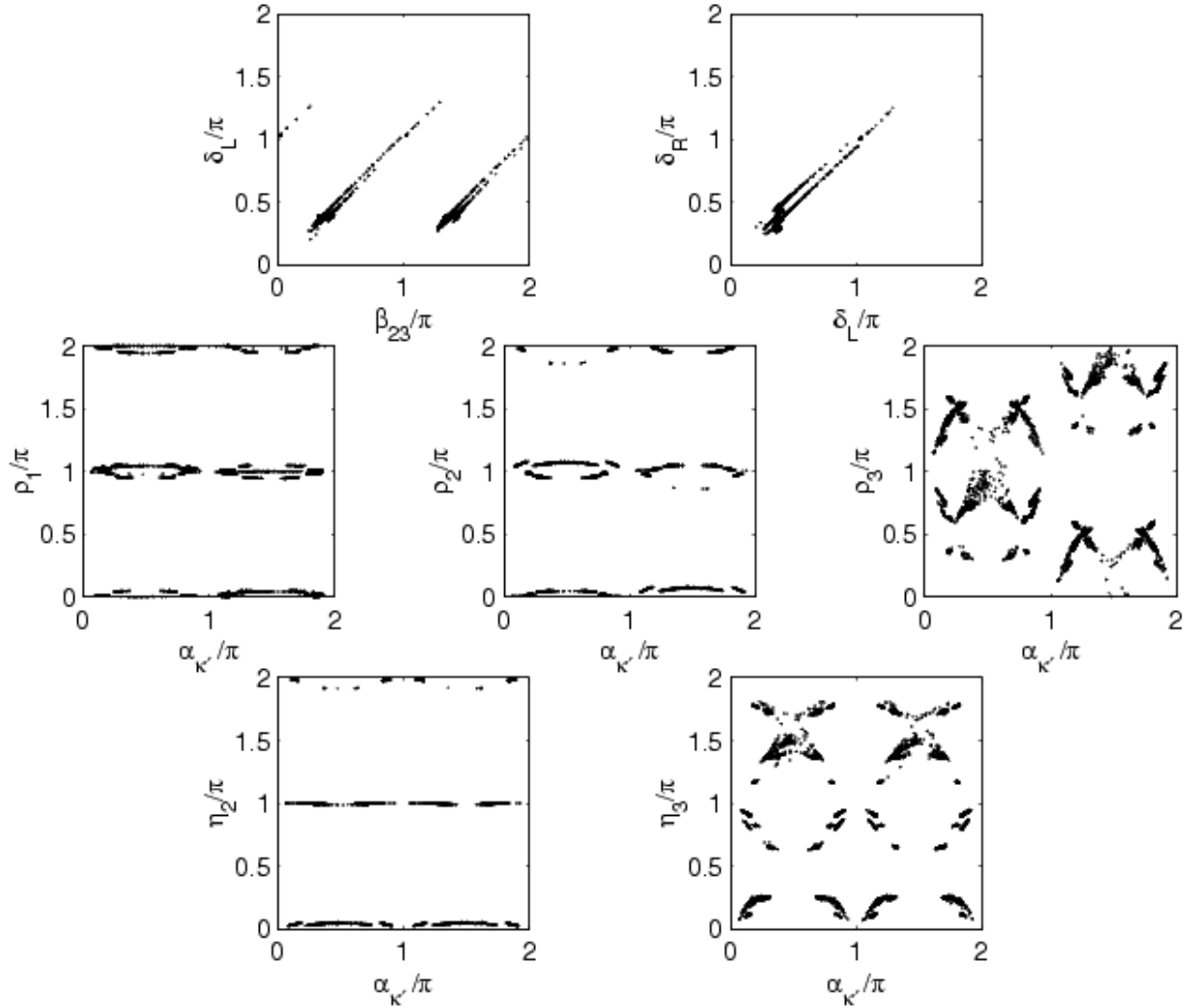}}
\caption{Left- and right-handed CKM phases for Level II solutions in the variable mass case.  The
data shown correspond to the mass range shown in the upper-right plot in Fig.~\ref{fig:solnsLevIIvar}.}
\label{fig:phasesvar}
\end{figure}
\begin{figure}
\resizebox{\textwidth}{!}{\includegraphics{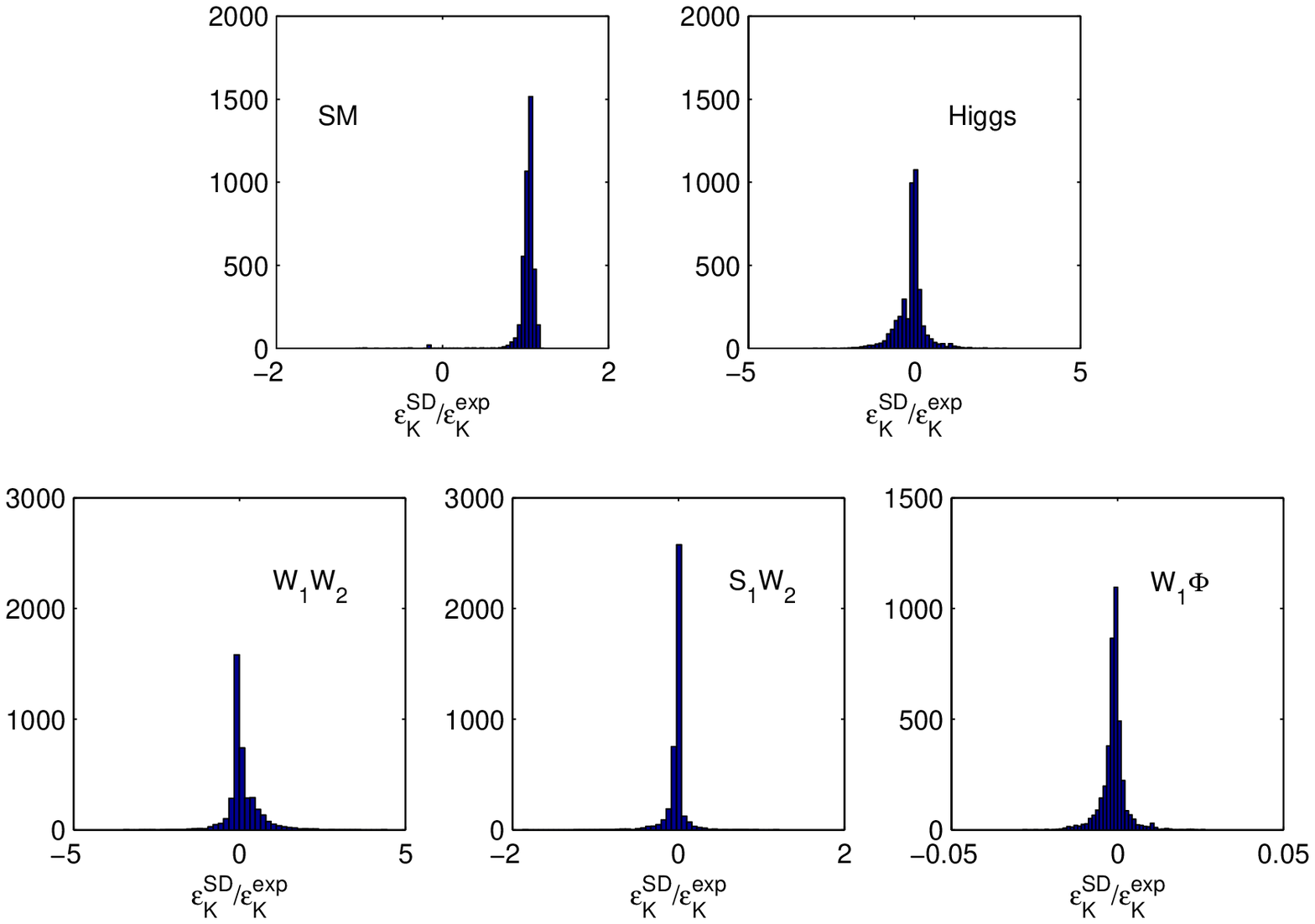}}
\caption{Frequency plots of the five short-distance contributions to $\epsilon_K$ for the variable
mass case.}
\label{fig:epskhistvar}
\end{figure}

Figures~\ref{fig:solnsLevIIvar}-\ref{fig:epskhistvar} show the results obtained when
$M_2$ and $M_H$ are allowed to vary over prescribed ranges.  Level II
solutions were found for Higgs masses as low as about 7~TeV, despite the apparently dangerous
tree-level FCNH contribution in Eq.~(\ref{eq:M12delmk}).  Solutions were also found
for $W_2$ masses as low as about 1.5 to 2~TeV.  In choosing the masses we have employed the
restrictions $M_H> M_2$, $M_2>1.4$~ TeV and $M_H>5$~TeV (otherwise, approximations in some of 
our theoretical expressions begin to lose some accuracy) and $M_2<13 M_H$ 
(perturbativity bound~\cite{olness2}).  We have also placed (somewhat arbitrary) upper limits 
on the $W_2$  and Higgs masses, as is evident in Fig.~\ref{fig:solnsLevIIvar}.

There are three sets of plots in Fig.~\ref{fig:solnsLevIIvar}.  The top pair shows the entire
range of Higgs and $W_2$ masses considered (on the right)
and the values of $\alpha_{\kappa^\prime}$ and $\beta_{23}$ for which solutions were found (on the left).
It would appear from this plot that the quasimanifest limit (real Higgs VEVs; $\alpha_{\kappa^\prime}=n\pi$) is disfavoured, at least
for the range of masses considered in the plot.
(The quasimanifest case can actually yield solutions in the decoupling limit, as we shall discuss below.)
The middle pair of plots shows the case in which $M_H<20$~TeV and indicates that for these ``moderate'' Higgs masses
the solutions in the $\beta_{23}$-$\alpha_{\kappa^\prime}$ plane form roughly horizontal bands.  The pseudomanifest case
(real $F$ and $G$; $\beta_{23}=n\pi$)
does not appear to be ruled out, but does seem at least to be slightly disfavoured.
(Recall that we could in principle come to different conclusions than Ball et al., since we do not
use the $\epsilon^\prime$ constraint in our analysis.)  The bottom pair of plots in Fig.~\ref{fig:solnsLevIIvar}
shows the approach to the ``decoupling limit'' ($M_H,M_2\to\infty$).  In the decoupling limit the
non-standard contributions all tend to zero, leaving only the SM contribution.  In contrast with the SB-LR,
our model survives into this limit, since the usual left-handed phase $\delta_L$ can be made to take on
any value in our model by choosing a suitable value for $\beta_{23}$
(see Eq.~(\ref{eq:deltal}) and Fig.~\ref{fig:LRphases}), whereas $\delta_L$ is quite close to either $0$ or
$\pi$ in the SB-LR~\footnote{Note that
among the two fundamental CP-violating phases
$\alpha_{\kappa^\prime}$ and $\beta_{23}$, the contribution from $\alpha_{\kappa^\prime}$ to the CKM phase $\delta_L$
is Cabibbo-suppressed by $\lambda=0.22$ relative to that from the phase $\beta_{23}$.
Thus models with vanishing $\beta_{23}$ (such as the SB-LR) may not
generate enough CP violation in general, and certainly not when the scale of new
physics is very high (i.e., in the decoupling limit).}.
Some additional investigation into the decoupling limit has shown that approximately vertical bands 
of solutions develop in the $\beta_{23}$-$\alpha_{\kappa^\prime}$ plane in this limit.
These bands are located near $\beta_{23}/\pi\simeq 0.4$ and $\beta_{23}/\pi\simeq 1.4$.
(These values for $\beta_{23}$ yield $\delta_L\simeq 1.2$~rad, which is just the usual result in the SM.)  
The beginnings
of these vertical ``decoupling limit bands'' are evident in Fig.~\ref{fig:solnsLevIIvar}.
Note that the decoupling limit version of Fig.~\ref{fig:LRphases} has very tightly constrained regions for
$\delta_L$ and also for $\delta_R$ (the constraint on $\delta_R$ is due to the strong correlation between $\delta_R$ and $\delta_L$).

Figure~\ref{fig:phasesvar} shows the left- and right-handed phases in the variable mass case.  Comparing with Fig.~\ref{fig:LRphases},
we see that the Level II constraints rule out many of the possible values that the phases could in principle assume.
The restrictions are not as severe, however, as in the case considered in Fig.~\ref{fig:LRphasesLevII510}, 
where the $W_2$ and Higgs masses were fixed to be 5 and 10~TeV, respectively.
Figure~\ref{fig:epskhistvar} displays the five short-distance contributions to $\epsilon_K$ 
(as fractions of $\epsilon_K^\textrm{\scriptsize exp}$) in 
the variable mass case.  Since many of the solutions actually correspond to quite large masses,
the non-standard distributions are all peaked around zero and the SM constribution
is peaked around unity.  Nevertheless, the data set includes cases 
in which the FCNH and $W_1$-$W_2$ contributions are relatively large (and typically of opposite sign).

\section{Discussion and Conclusions}
\label{sec:concl}

The Left-Right Model provides a viable and aesthetically pleasing extension of the SM.  We have presented
a relatively exhaustive numerical investigation of a nonmanifest ``top-inspired'' version of the Left-Right Model in which the
Higgs VEVs are taken to be in the ratio $m_b:m_t$.  This version of the model is very attractive 
in that it quite naturally reproduces the quark mass
and rotation angle hierarchies.  It has often been the case in the past that 
studies of the nonmanifest Left-Right Model have relied on various ansatzes regarding the form of the 
right-handed CKM matrix.  In the present work we have solved for this matrix numerically.  Our numerical
work has yielded the intriguing result that the right-handed rotation angles and
the phase $\delta_R$
are very similar in size to their left-handed counterparts.
These relations have been corroborated analytically in Appendix~\ref{sec:appendixb}.
One interesting feature of the model is that, unlike the SB-LR,
it reduces to the SM in the decoupling limit.

One of the key insights in the present work is that 
unitary rotations may be used to rotate away many superfluous degrees of freedom in the quark Yukawa
matrices $F$ and $G$, yielding mass matrices that contain only
two fundamental CP-odd phases.  CP-violating quantities such as $\epsilon_K$ and 
$\sin 2\beta_{CKM}^\textrm{\scriptsize eff}$ may then be used to place constraints on these two phases.
Our numerical study indicates that the combined consideration of the neutral $K$ and
$B$ systems leads to quite a strong reduction in the size of the available parameter space.
In particular, the two CP-odd phases $\alpha_{\kappa^\prime}$ and $\beta_{23}$ are confined to rather small regions.  From the
vantage point of this numerical investigation, and with the
range of masses considered here, it would appear
that both the quasimanifest (real Higgs VEVs) and pseudomanifest (real Yukawa couplings)
versions of the model are disfavoured.  The latter of these results is in agreement with recent work by 
Ball et al.~\cite{ball1}, although some of our numerical results appear to be mildly different from theirs.
One very intriguing result of the present work is that $W_2$ and Higgs masses as light as about 2~TeV and 7~TeV, 
respectively, are not inconsistent
with current experimental constraints.

\begin{acknowledgments}
We would like to thank Jeremy Case for helpful conversations and Jason O'Kane
and Carl Daudt for technical assistance.
This research was supported in part by the U.S. Department of Energy
contract numbers  DE-AC02-98CH10886 (BNL) and
DE-FG03-96ER40969 (Oregon).
K.K. and J.K. were supported by an award from Research Corporation and 
J.L. was supported by the SRTP at Taylor University.  
G.W. and K.K. would like to thank the BNL high energy theory group for its hospitality 
during the initial stages of this work. 
\end{acknowledgments}

\appendix
\section{Adaptive Monte Carlo Algorithm}
\label{sec:montecarlo}

The numerical solution of our model has been accomplished using an adaptive
Monte Carlo algorithm.  The goal of the algorithm is to find sets of input
parameters $(f_{ii},g_{ij},\ldots)$ that satisfy the Level I or Level II
constraints to within some required tolerance.  
Table~\ref{tab:chi2} lists the various constraints employed and
Eqs.~(\ref{eq:conLevI}) and (\ref{eq:conLevII}) give the definitions
of $\chi^2$ for Level I and II constraints, respectively.
The basic procedure is to generate random values for the various input parameters
(see Table~\ref{tab:inputranges})
and then to ``zoom in'' on a solution by searching for small values of $\chi^2$.
The procedure consists of $N_\textrm{\scriptsize it}$ iterations (or ``zooms''),
with each iteration consisting of the construction and diagonalization of 
$N_\textrm{\scriptsize calc}$ separate sets of mass matrices.
For each successive iteration, the sizes of the input parameter
ranges are reduced and are centered on the ``best'' (as determined by $\chi^2$)
set of input parameters for the run so far.
For a typical run, $N_\textrm{\scriptsize calc}\sim 15$ and
$N_\textrm{\scriptsize it}\sim 5000$.

The specific procedures for generating Level I and II solutions differ slightly;
these differences are explained below after the description of the general algorithm.
The general algorithm for finding a solution is as follows:
\begin{enumerate}
\item For the $n^\textrm{\scriptsize th}$ iteration ($n=0, 1, 2,\ldots N_\textrm{\scriptsize it}$),
randomly generate a set of values for the relevant input parameters $x_i$
(see Table~\ref{tab:inputranges})
within the ranges
\begin{eqnarray}
	x_{i,n}^\textrm{\scriptsize cent} \pm \xi_n \Delta_i \; ,
\end{eqnarray}
where
\begin{eqnarray}
	\xi_n = \left\{\begin{array}{ll}
		1, & n=0 \\
		(1.5 n)^{-1}, & n>0 \; .\\
		\end{array}\right.
		\label{eq:xi1}
\end{eqnarray}
The above functional form for $\xi_n$ was found to be convenient and relatively
efficient.  One could in principle choose a form for $\xi_n$ that decreases more quickly
(exponentially, say), but we were not successful in getting such
forms to converge to solutions.~\footnote{Since our routine allows the $f_{ii}$ and $g_{ij}$
to ``wander'' out of their original ranges, extra precautions were taken to ensure
that $f_{33}$, $g_{12}$, $g_{13}$ and $g_{23}$ all remained positive.}
\item Numerically diagonalize the mass matrices (see Sec.~\ref{sec:qmm}) and
determine the quark masses and the right- and left-handed CKM matrices.
\item Evaluate the appropriate $\chi^2$ (either $\chi_{(I)}^2$
or $\chi_{(II)}^2$) for the set of input parameters (see Eqs.~(\ref{eq:conLevI})
and (\ref{eq:conLevII})).
\item Return to Step 1 and repeat the process $N_\textrm{\scriptsize calc}$ times.  
Keep track of the set of input parameters that has yielded
the lowest value of $\chi^2$ for the run so far, calling this set $\{x_{i}^\textrm{\scriptsize best}\}$.
\item After repeating the process $N_\textrm{\scriptsize calc}$ times, 
set the ``central values'' of the various input parameters for the next iteration
to the values that have yielded the best $\chi^2$ so far for the run:
\begin{eqnarray}
	x_{i,n+1}^\textrm{\scriptsize cent} =  x_{i}^\textrm{\scriptsize best}\; .
\end{eqnarray}
Increment $n$ by one and return to Step 1.
Repeat the entire process $N_\textrm{\scriptsize it}$ times.  Note that 
on returning to Step 1, $n$ has increased, so $\xi_n$ has been reduced in size.
\item Check the individual $\chi_i^2$ once all  $N_\textrm{\scriptsize it}$ iterations are completed.  
If each of the $\chi_i^2\le 1$, the input parameter set is a solution.
\end{enumerate}

A few modifications were made to the above procedure in order to increase its efficiency.
For example, it was found empirically that $\chi^2$ decreased rather quickly
on runs that actually resulted in a solution.  We thus modified the procedure
so that runs were abandoned if $\chi^2$ had not decreased below some threshold value
after a specified number of iterations.

\subsection{Level I Solutions}

As noted in Sec.~\ref{sec:LevI}, the six quark masses and three rotation angles provide nine essential constraints
on the input parameter space.  As a result, Level I solutions may
be found for any pair of phases $\alpha_{\kappa^\prime}$ and $\beta_{23}$ (the nine constraints
simply act to constrain the nine input parameters $f_{ii}$ and $g_{ij}$).  In
fact, in the limit that $\chi_{(I)}^2\rightarrow 0$, there appear to be
32 solutions for each pair of values of $\alpha_{\kappa^\prime}$ and $\beta_{23}$.  In
searching for a Level I solution, only the $f_{ii}$ and $g_{ij}$ are ``zoomed in upon'';
$\alpha_{\kappa^\prime}$ and $\beta_{23}$ are fixed at the beginning of a particular run and are not
altered throughout the course of the run.  The Level I solutions for
Figs.~\ref{fig:FGspace}-\ref{fig:rotnangles} were generated by setting $N_\textrm{\scriptsize calc}=15$ and
$N_\textrm{\scriptsize it}=1700$.  Slightly modified procedures were used to generate
the data for Figs.~\ref{fig:delmb} and \ref{fig:epsksin2b}.

\subsection{Level II Solutions}

In order to find a Level II solution, one must specify
values for $M_2$ and $M_H$ (since $\epsilon_K$, etc., depend on these).  
As Level II solutions involve the addition of several new constraints
compared to Level I solutions,
it is convenient to allow both $\alpha_{\kappa^\prime}$ and $\beta_{23}$ to
be free parameters while zooming in on a solution.  $M_2$ and $M_H$, however, 
may remain fixed for any particular run.

A few slight modifications must be made to the general algorithm
when searching for a Level II solution, since a straightforward application
of the algorithm does not seem to lead to solutions.  The reason for the problem appears to
be the inclusion of $\epsilon_K$ in the evaluation of $\chi_{(II)}^2$.
$\epsilon_K$ is strongly suppressed in the SM due in part to the presence of small CKM matrix
elements~\cite{bbl}.  At the beginning of a particular run, $\epsilon_K$ would typically be orders of
magnitude too large because the elements in the CKM matrix would not initially have the correct hierarchy.
A similar problem
occurs for the right-handed contributions to $\epsilon_K$.  
The massive deviation of $\epsilon_K$ from its experimental value at the beginning of a run leads to a very large
contribution to $\chi^2$ and upsets the zooming process.
In order to get around the
problem, we substitute approximate values (close to the known experimental values for the left-handed angles)
for both the left- and right-handed rotation angles when determining the contribution of 
$\epsilon_K$ (and also of $\sin 2\beta_{CKM}^\textrm{\scriptsize eff}$)
to $\chi_{(II)}^2$ for the first several hundred iterations.  At some point in each run
a switch is made such that the true numerical versions of the left- and right-handed CKM matrices 
are used.  (Note that part of the reason for the success of this trick
is the relatively good agreement between the left- and right-handed rotation angles evident
in Fig.~\ref{fig:rotnangles}.)  The Level II solutions for Figs.~\ref{fig:fgspaceLevII510}-\ref{fig:epskhistvar}
were generated by setting $N_\textrm{\scriptsize calc}=18$ and $N_\textrm{\scriptsize it}=8000$.

\section{Appendix: Rotation angle and phase relations in the model}
\label{sec:appendixb}

In this appendix, we derive analytical relations between the left- and 
right-handed rotation angles and the CKM phases $\delta_L$ and $\delta_R$
for the model considered in this paper.  The analysis is greatly simplified
by the smallness of the ratio of the VEV's: $|\kappa^\prime/\kappa| = m_b/m_t$.

\subsection{Angle relations}

      As discussed in Sec.~\ref{sec:model},
the quark mass matrices in this model exhibit a surprising 
simplicity due to the hierarchy of the VEVs of the scalar bidoublet.
In particular, ${\mathcal M}_u  =  \kappa F + \kappa^{\prime *} G$
is nearly diagonal in our choice of basis, and its small rotation angles
can be safely neglected compared to the corresponding CKM rotations;
i.e.,
\begin{equation}
  |V_L^U| \simeq  |V_R^U| \simeq \textrm{diag(1, 1, 1)} .
\end{equation} 
Thus, both left-handed (LH) CKM rotations and right-handed (RH) rotations arise
solely from ${\mathcal M}_d  =  \kappa^\prime F + \kappa^* G$.
Note that ${\mathcal M}_d$ is neither Hermitian (due to the phase $\alpha_{\kappa^\prime}$)
nor symmetric (due to the phase $\beta_{23}$), and we need two separate unitary
rotation matrices $V_L^D$ and $V_R^D$ to diagonalize ${\mathcal M}_d$.
We will show in this section that the LH and RH rotation angles are
closely related in this model due to the hierarchical structure
in the observed quark mass spectrum and in the CKM angles.
This feature of the model is evident from the numerical results presented
in the text.

   We start by noting the approximate but useful hierarchies, 
$m_u : m_c : m_t \sim \lambda^8 : \lambda^4 : 1$,
$m_d : m_s : m_b \sim \lambda^4 : \lambda^2 : 1$, and
$V_{ub} \sim \lambda^4$, $V_{cb} \sim \lambda^2$,
$V_{us} = \lambda = 0.22$.
${\mathcal M}_d {\mathcal M}_d^{\dagger}$ is determined
to a good approximation from the LH CKM matrix, 
with the order of magnitude
of the different matrix elements given by,
\begin{equation} \label{eq:hier}
\left| {\mathcal M}_d {\mathcal M}_d^{\dagger} \right|
 \sim m_b^2 \left( \begin{array}{ccc}
             \lambda^6 & \lambda^5 &  \lambda^4 \\
                    \lambda^5 & \lambda^4  &  \lambda^2 \\
                    \lambda^4 & \lambda^2 &  1 \\
             \end{array}\right) \! .
\end{equation}
The hierarchical structure of this matrix will be useful
as we examine the rotation angles.

  The matrix ${\cal M}_d$ can be rewritten as
\begin{equation}
 {\mathcal M}_d = H + P \simeq H + m e^{i\alpha_{\kappa^\prime}} \left( \begin{array}{ccc}
                                          0 & 0 & 0 \\
                                          0 & 0 & 0 \\
                                          0 & 0 & 1 \\
                         \end{array}\right) \! ,
\end{equation}
where $H = \kappa^* G$ is Hermitian, $m = |\kappa^\prime f_{33}| \sim m_b$,
and $P = \kappa^\prime F$ and has 
been approximated by neglecting the small (1,1) and (2,2) elements 
for simplicity of analysis. The inclusion of all 3 diagonal elements of $P$
is straightforward and does not affect our result.
The LH and RH rotation matrices can be separately determined
from ${\mathcal M}_d {\mathcal M}_d^{\dagger}$ and
${\mathcal M}_d^{\dagger} {\mathcal M}_d$, respectively.
\begin{eqnarray}
  {\mathcal M}_d {\mathcal M}_d^{\dagger} & = &
     (H + P)(H + P^*)                  \nonumber \\
  & \simeq &  
  \left( \begin{array}{ccc}
     H^2_{11}  & H^2_{12}  & H^2_{13} + e^{-i\alpha_{\kappa^\prime}} H_{13} m \\
     H^2_{21}  & H^2_{22}  & H^2_{23} + e^{-i\alpha_{\kappa^\prime}} H_{23} m \\
     H^2_{31} + e^{i\alpha_{\kappa^\prime}} H_{31} m
       & H^2_{32} +  e^{i\alpha_{\kappa^\prime}} H_{32} m
     & H^2_{33} + m^2 + 2\cos\alpha_{\kappa^\prime} H_{33} m \\
     \end{array}\right) \! , \\
{\mathcal M}_d^{\dagger} {\mathcal M}_d & = & (H + P^*)(H + P) =
  {\mathcal M}_d {\mathcal M}_d^{\dagger} ( \alpha_{\kappa^\prime} \rightarrow -\alpha_{\kappa^\prime})
\end{eqnarray}
where $H^2_{ij}$ denotes the $(i,j)$ element of $H^2$. 
Therefore, the LH and RH rotation angles are also related
by $\alpha_{\kappa^\prime} \rightarrow -\alpha_{\kappa^\prime}$, and for this reason, one would expect
them to be of the same order of magnitude; i.e.,
\begin{equation}
\theta^L_{ij} / \theta^R_{ij} = {\cal O}(1) \, .
\end{equation}
We thus arrive at the hierarchical structure of the 
${\mathcal M}_d$ (thus $H$) matrix,
\begin{equation}  \label{eq:Md}
|{\mathcal M}_d | \sim 
  m_b \left( \begin{array}{ccc}
             \lambda^4 & \lambda^3 &  \lambda^4 \\
                    \lambda^3 & \lambda^2  &  \lambda^2 \\
                    \lambda^4 & \lambda^2 &  1 \\
             \end{array}\right) 
        \sim |H|
\, ,
\end{equation}
where only the order of magnitude of each matrix element is given.
To be more precise,  $0 \le |H_{33}|/m_b \le {\cal O}(1)$, and this is 
because $|P_{33}| = |\kappa^\prime f_{33}| \sim m_b$.
We have checked that Eqs.~(\ref{eq:theta23}), (\ref{eq:theta13}), and 
(\ref{eq:theta12}) are valid even when $|H_{33}|/m_b \ll 1$.

 Due to the hierarchical structure of ${\mathcal M}_d {\mathcal M}_d^{\dagger}$
(see Eq.~(\ref{eq:hier})), the LH angle $\theta^L_{23}$ (i.e., $|V_{cb}|$) is 
simply given by~\cite{kuo}
\begin{equation} \label{eq:23}
  \theta^L_{23} \simeq \left|
         \frac{ ({\mathcal M}_d {\mathcal M}_d^{\dagger})_{23} }{
         ({\mathcal M}_d {\mathcal M}_d^{\dagger})_{33} } \right|
    = \left| \frac{H_{23} H_{33} + H_{22} H_{23} + H_{21} H_{13} 
       + e^{-i\alpha_{\kappa^\prime}} H_{23} m} 
         {({\mathcal M}_d {\mathcal M}_d^{\dagger})_{33} }
            \right| \! .
\end{equation}
 It is easy to see that $({\mathcal M}_d {\mathcal M}_d^{\dagger})_{33}$
is even under $\alpha_{\kappa^\prime} \rightarrow -\alpha_{\kappa^\prime}$.
In the numerator in Eq.~(\ref{eq:23}),  only $H_{23}$ and $e^{-i\alpha_{\kappa^\prime}}$ 
are complex. If we ignore the small term of $H_{21} H_{13}$,
we can then factor out $H_{23}$ and observe that the numerator (thus 
$\theta^L_{23}$) is invariant under $\alpha_{\kappa^\prime} \rightarrow -\alpha_{\kappa^\prime}$.
Thus $ \theta^R_{23} = \theta^L_{23} (\alpha_{\kappa^\prime} \rightarrow -\alpha_{\kappa^\prime}) 
\simeq \theta^L_{23}$. 
The inclusion of the small term $H_{21} H_{13}$ introduces a tiny
correction to this equality relation,
\begin{equation} \label{eq:theta23}
    \theta^R_{23} =  \theta^L_{23} \times ( 1 + {\cal O}(\lambda^5)) \, ,
\end{equation}
where we have made use of Eq.~(\ref{eq:Md}). 
This correction is of order $0.1\%$ and is in good agreement with 
our numerical analysis (see the second plot in Fig.~\ref{fig:rotnangles}).  

    From Eq.~(\ref{eq:hier}), one can reason that $\theta^L_{13}$ is
given by
\begin{equation} \label{eq:13}
  \theta^L_{13} \simeq \left|
         \frac{ ({\mathcal M}_d {\mathcal M}_d^{\dagger})_{13} }{
         ({\mathcal M}_d {\mathcal M}_d^{\dagger})_{33} } \right|
   =  \left| \frac{ e^{-i\alpha_{\kappa^\prime}} H_{13} m +
          H_{13} H_{33} + H_{12} H_{23} + H_{11}H_{13} } 
        { ({\mathcal M}_d {\mathcal M}_d^{\dagger})_{33} }
            \right| \, .
\end{equation}
Substituting the different  
terms with their orders-of-magnitude and phases, we have
\begin{equation} 
\theta^L_{13} 
  =   \left|e^{-i\alpha_{\kappa^\prime}} + {\cal O}(1) + e^{-i\beta_{23}} {\cal O}(\lambda)
     \right| \times {\cal O}(\lambda^4) \, , 
\end{equation} 
which gives the right size for $|V_{ub}|$.
The corresponding RH angle can then be deduced 
from $\theta^L_{13}$ with $\alpha_{\kappa^\prime} \rightarrow -\alpha_{\kappa^\prime}$, and we get
\begin{equation} \label{eq:theta13}
    \theta^R_{13} =  \theta^L_{13} \times ( 1 + {\cal O}(\lambda)) \, .
\end{equation}
This ${\cal O}(\lambda)$ correction well explains the $20\%$ fluctuation
around unity in the third plot of Fig.~\ref{fig:rotnangles}.
  
      The last step in the diagonalization of 
${\mathcal M}_d {\mathcal M}_d^{\dagger}$ involves a (1,2) rotation.
To first approximation, the matrix elements of the (1,2) submatrix
are invariant under $\alpha_{\kappa^\prime} \rightarrow -\alpha_{\kappa^\prime}$, and we get
\begin{equation}
\theta^R_{12} \simeq  \theta^L_{12} \, .
\end{equation}
To find out the correction to this relation,
we need to include the ``residual effect" on the $(1,2)$ and $(2,2)$ elements of
${\mathcal M}_d {\mathcal M}_d^{\dagger}$ from the $(2,3)$ rotation. 
The $(1,2)$ element is modified as
\begin{eqnarray*}
H^2_{12} & \longrightarrow &
H^2_{12} + (H^2_{32} + e^{i\alpha_{\kappa^\prime}} H_{32} m ) 
                 ( H^2_{13} + e^{-i\alpha_{\kappa^\prime}} H_{13} m ) /{\cal O}(m_b^2) \\
        & = & 
         \left\{ 
         1  + {\cal O}(\lambda) e^{-i\beta_{23}} 
              \left[e^{i\alpha_{\kappa^\prime}} + e^{-i\alpha_{\kappa^\prime}} + {\cal O}(1) \right]
                 + {\cal O}(\lambda^2) \left[e^{i\alpha_{\kappa^\prime}} + {\cal O}(1) \right] 
            \right\} 
             \times {\cal O}(\lambda^5) m_b^2 
             \, . 
\end{eqnarray*} 
Interestingly, the ${\cal O}(\lambda)$ term is invariant under 
$\alpha_{\kappa^\prime} \rightarrow -\alpha_{\kappa^\prime}$, and the leading non-invariant term
appears at a higher order of $\lambda^2$. 

   One can similarly calculate the effect of the $(2,3)$ rotation on the
$(2,2)$ element of ${\mathcal M}_d {\mathcal M}_d^{\dagger}$.
We note that the non-invariant term under $\alpha_{\kappa^\prime} \rightarrow -\alpha_{\kappa^\prime}$
is of ${\cal O}(\lambda^5)$ relative to the invariant term.
The LH rotation angle $\theta^L_{12}$ can now be calculated, and is given 
to first approximation by the ratio of the modified $(1,2)$ and $(2,2)$
 elements. The RH angle $\theta^R_{12}$ can be obtained from $\theta^L_{12}$
with the substitution $\alpha_{\kappa^\prime} \rightarrow -\alpha_{\kappa^\prime}$.  We thus get
\begin{equation} \label{eq:theta12}
    \theta^R_{12} =  \theta^L_{12} \times ( 1 + {\cal O}(\lambda^2)) \, .
\end{equation}
 The ${\cal O}(\lambda^2)$ correction nicely explains the $\sim 4\%$ deviation
from unity as presented in the first plot of Fig.~\ref{fig:rotnangles}.

    Note that when either of the two phases vanishes, we have the exact relations
$\theta^L_{ij} = \theta^R_{ij}$.  In particular, when 
$\alpha_{\kappa^\prime} = 0 ~{\rm or}~ \pi$ (quasimanifest case),
${\mathcal M}_d {\mathcal M}_d^{\dagger} = 
{\mathcal M}_d^{\dagger}{\mathcal M}_d$ because $P = P^*$. This yields
identical LH and RH rotation angles.
On the other hand, when $\beta_{23} = 0 ~{\rm or}~ \pi$ (pseudomanifest case),
it is easily seen that the expressions for $\theta^L_{ij}$ are invariant
under $\alpha_{\kappa^\prime} \rightarrow -\alpha_{\kappa^\prime}$, thus
$\theta^L_{ij} = \theta^R_{ij}$.

\subsection{Phase relations}

   Due to the hierarchical structure of the quark mass matrices in the model
considered in this paper, 
 we can use the triangular matrix technique
developed in Ref.~\onlinecite{kuo} to solve for the CKM phases $\delta_L$ and
$\delta_R$.  In the triangular form, each mass matrix element has a simple
correspondence with a quark mass or a rotation angle, and the CKM 
phases $\delta_L$ and $\delta_R$ are equal to linear combinations of the phases of certain 
elements of the up- and down-type quark mass matrices~\cite{kuo}.  
For practical purposes, ${\mathcal M}_u$ can be considered as diagonal
in our model, and the CKM phase $\delta_L$ ($\delta_R$) depends on
the phases of four matrix elements of ${\mathcal M}_d$ rewritten in the upper 
(lower) triangular form~\cite{kuo}.           

      For the purpose of comparing $\delta_L$ and $\delta_R$ later, 
Eq.~(\ref{eq:Md}) can be rewritten with real and ${\cal O}(1)$ coefficients
as follows,
\begin{equation}  \label{eq:Md2}
{\mathcal M}_d  \simeq 
  m_b \left( \begin{array}{ccc}
    a_1 \lambda^4 & a_2 \lambda^3 &  a_3 \lambda^4 \\
    a_2 \lambda^3 & b_1 \lambda^2 
      e^{i d \lambda \sin \alpha_{\kappa^\prime}}
      & b_2 \lambda^2 e^{i\beta_{23}} \\
    a_3  \lambda^4 & b_2 \lambda^2 e^{-i\beta_{23}} & e^{i\alpha^\prime}   \\
             \end{array}\right) 
\, .
\end{equation}
Note that $m_b e^{i\alpha^\prime} \equiv \kappa^\prime f_{33} +
     \kappa^* g_{33}$, and that we have used $\alpha^\prime$ to distinguish 
from $\alpha_{\kappa^\prime} = \arg(\kappa^\prime)$.
The ${\cal O}(\lambda)$ phase factor in ${\mathcal M}_d(2,2)$ 
comes from $\kappa^\prime f_{22}$, and 
$d \simeq f_{22} |\kappa^\prime| /(m_b \lambda^3 b_1)$ is an ${\cal O}(1)$
coefficient. 
On the other hand, $\kappa^\prime f_{11}$ modifies 
 ${\mathcal M}_d(1,1)$ by  $1+{\cal O}(\lambda^4)$ and is thus neglected. 
For the phase relations, we will include corrections up to 
${\cal O}(\lambda)$.

   Consider first the phase $\delta_L$ in the LH CKM matrix.  To this end,
we apply RH rotations to Eq.~(\ref{eq:Md2}) to convert it to upper-triangular
form,
\begin{equation}  \label{eq:upper}
{\mathcal M}_d  \rightarrow {\mathcal M}_d^A \simeq  
  m_b
\left( \begin{array}{ccc}
  a_1^\prime \lambda^4 & a_2 \lambda^3 &  
       a_3 \lambda^4 \left[ 1 + c \lambda e^{i(\beta_{23}+\alpha^\prime)}
                     \right] \\
   0  & b_1 \lambda^2 e^{i d \lambda \sin \alpha_{\kappa^\prime}}
 & b_2 \lambda^2 e^{i\beta_{23}} \\
   0 &  0  & e^{i\alpha^\prime}   \\
             \end{array}\right) 
\, ,
\end{equation}
where $c = a_2b_2/a_3$,  $a_1$ has changed to $a_1^\prime$ which itself
carries an ${\cal O}(\lambda)$ phase, and 
higher-order corrections in $\lambda$ are ignored.

   The CKM phase $\delta_L$ can now be expressed in terms of four elements
of ${\mathcal M}_d^A$~\cite{kuo},
\begin{equation}  \label{eq:delL}
\delta_L \simeq \arg\left[ \frac{{\mathcal M}_d^A(1,2) {\mathcal M}_d^A(2,3)}
                                {{\mathcal M}_d^A(1,3) {\mathcal M}_d^A(2,2)}
                     \right]
  \simeq \beta_{23} -  c \lambda \sin(\beta_{23}+\alpha^\prime) 
   - d \lambda \sin \alpha_{\kappa^\prime}
+ n \pi \, ,
\end{equation}
 where $n\pi=\arg\left(\frac{a_2b_2}{a_3b_1}\right)$ and can be $0$ or 
$\pm \pi$.

     To obtain an expression for $\delta_R$ in the RH CKM matrix, we apply 
LH rotations to Eq.~(\ref{eq:Md2}) to convert it to lower-triangular
form,
\begin{equation}  \label{eq:lower}
{\mathcal M}_d  \rightarrow {\mathcal M}_d^B \simeq
  m_b
\left( \begin{array}{ccc}
  a_1^\prime \lambda^4 & 0 & 0 \\    
   a_2 \lambda^3  & b_1 \lambda^2 e^{i d \lambda \sin \alpha_{\kappa^\prime}} 
  & 0 \\
   a_3 \lambda^4 \left[ 1 + c \lambda e^{-i(\beta_{23}-\alpha^\prime)} 
                     \right]  &  
        b_2 \lambda^2 e^{-i\beta_{23}}  & e^{i\alpha^\prime}   \\
             \end{array}\right) 
\, .
\end{equation}
 The phase $\delta_R$ can be similarly obtained,
\begin{equation}  \label{eq:delR}
\delta_R \simeq \arg\left[ \frac{{\mathcal M}_d^B(2,2) {\mathcal M}_d^B(3,1)}
                                {{\mathcal M}_d^B(2,1) {\mathcal M}_d^B(3,2)}
                     \right]
  \simeq \beta_{23} -  c \lambda \sin(\beta_{23}-\alpha^\prime) 
 + d \lambda \sin \alpha_{\kappa^\prime}
+ n \pi \, ,
\end{equation}
where $n\pi= \arg\left(\frac{a_3b_1}{a_2b_2}\right) = 
\arg\left(\frac{a_2b_2}{a_3b_1}\right)$.

  Some discussion is in order regarding several features and limits of the model.
\begin{enumerate}
\item \textit{$\delta_R \approx \delta_L$.}
 Comparing the expressions for $\delta_L$ and $\delta_R$, we see that
they differ only in the sign of $\alpha_{\kappa^\prime}$ 
(thus $\alpha^\prime$), as noted when we examined
the LH and RH angles.
More specifically,
\begin{equation}  \label{eq:delLR}
\delta_R = \delta_L + 2  c \lambda \cos\beta_{23} \sin \alpha^\prime 
             + 2 d \lambda\sin \alpha_{\kappa^\prime}
             + {\cal O}(\lambda^2) \, .
\end{equation}
Therefore, $\delta_R$ and $\delta_L$ become degenerate if 
$\alpha_{\kappa^\prime} = 0$ or $\pi$.
Noting that $c, d ={\cal O}(1)$ and $2\lambda = 0.44$, 
Eq.~(\ref{eq:delLR}) well explains our numerical relation of Eq.~(\ref{eq:deltaRapprox}) and
the second plot in Fig.~\ref{fig:LRphases},
\begin{eqnarray*}
\delta_R &\approx & \delta_L\pm 0.50~\textrm{rad} \, .
\end{eqnarray*}
 In other words, $\delta_L$ and $\delta_R$ are equal up to 
${\cal O}(\lambda)$ corrections from a non-zero $\alpha_{\kappa^\prime}$.

\item \textit{$\delta_L \approx \beta_{23} \approx \delta_R$ (mod $\pi$).}
As $c, d ={\cal O}(1)$ and $\lambda = 0.22$, we see that
Eq.~(\ref{eq:delL}) is in good agreement with our numerical relation 
of Eq.~(30) and the first plot in Fig.~\ref{fig:LRphases}:
\begin{eqnarray*}
\delta_L &\approx & \beta_{23} + n \pi \pm 0.25~\textrm{rad} \nonumber \, .
\end{eqnarray*}
A similar expression is valid for $\delta_R$.
In both cases, the CKM phase is simply equal to $\beta_{23}$ (mod $\pi$)
up to ${\cal O}(\lambda)$ corrections. In particular, the contribution
to $\delta_{L,R}$ from $\alpha_{\kappa^\prime}$ is Cabibbo-suppressed
by $\lambda$ relative to that from $\beta_{23}$.  

\item \textit{(Quasi)manifest limit: $\delta_L = \delta_R$.} 
 In the limit $\alpha_{\kappa^\prime} = 0 ~ {\rm or} ~\pi$ 
(thus $\alpha^\prime = 0 ~ {\rm or} ~\pi$), we recover the (quasi)manifest 
Left-Right Symmetric Model, and
$\delta_L = \delta_R$ (exactly), as is well known for this scenario.
For our model, we have
\begin{equation}  \label{eq:same}
\delta_R = \delta_L \simeq \left\{ \begin{array}{l}
\beta_{23} -  c \lambda \sin\beta_{23} + n \pi 
 ~~~~~~~~~  (\alpha^\prime = 0)  \\
\beta_{23} +  c \lambda \sin\beta_{23} + n \pi
 ~~~~~~~~~  (\alpha^\prime = \pi) \, . 
\end{array} \right.
\end{equation}

\item \textit{Pseudomanifest limit (or SB-LR): $\delta_R = - \delta_L$
and $|\delta_{L,R} - n \pi|  \le  {\cal O}(\lambda)$.}
In the limit $\beta_{23} = 0 ~ {\rm or} ~\pi$, we have the pseudomanifest Left-Right Symmetric Model,
or the SB-LR.
As is well known for this case, we have the exact relation
$\delta_L = - \delta_R$.  In our model, we get
\begin{equation}  \label{eq:opposite}
\delta_R = - \delta_L \simeq \left\{ \begin{array}{ll}
c \lambda \sin \alpha^\prime + d \lambda \sin \alpha_{\kappa^\prime} + n \pi 
 ~~~~~~~~~& (\beta_{23} = 0) \\
-c \lambda \sin \alpha^\prime + d \lambda \sin \alpha_{\kappa^\prime}
    + (n+1) \pi
 ~~~~~~~~~ & (\beta_{23} = \pi) \, .
\end{array} \right.
\end{equation}
Interestingly, the magnitudes of both phases are Cabibbo-suppressed, i.e.
\begin{eqnarray}
|\delta_{L,R} - m \pi| & \le & {\cal O}(\lambda) 
~~~~~~~~~ (\beta_{23} = 0 ~ {\rm or} ~\pi) \, ,
\end{eqnarray}
where $m=0 ~ {\rm or} ~ 1$.
Note that this suppression arises before we impose
any CP-violating constraints on the model.  
Our analytical result is consistent with the numerical findings of 
Ball et al.~\cite{ball1},
\begin{eqnarray*}
  |\delta_L - m \pi| & \le & 0.25    ~~~~~~~(m = 0, 1) \, .
\end{eqnarray*}
This Cabibbo-suppression of the CKM phases may help explain why the SB-LR is
disfavoured by the 
$\sin 2\beta_{CKM}^\textrm{\scriptsize eff}$ measurement.
\end{enumerate}

\bibliography{paper}
\end{document}